\documentclass[%
 reprint,
 superscriptaddress,
 amsmath,amssymb,
 aps,
 pra,
floatfix,
]{revtex4-1}
\usepackage{amsmath}
\usepackage{amssymb}
\usepackage{graphicx}
\usepackage{color}
\usepackage{mathtools}
\usepackage{hyperref}
\usepackage[caption=false]{subfig}
\usepackage[colorinlistoftodos]{todonotes}
\usepackage{lipsum}
\usepackage{bbold}
\usepackage{tabularx}
\usepackage{float}
\usepackage{comment}

\usepackage{tikz}
\usepackage{circuitikz}
\usepackage{tikz-timing}
\usetikzlibrary{arrows.meta,decorations.pathmorphing,decorations.pathreplacing,positioning,shapes,fadings}

\newcommand{\ket}[1]{\left|#1\right\rangle}
\newcommand{\bra}[1]{\left\langle#1\right|}

\DeclareMathOperator{\iim}{Im}
\DeclareMathOperator{\Tr}{Tr}

\graphicspath{ {main_figures/} {supplement_figures/} }

\makeatletter
\@ifundefined{textcolor}{}
{%
 \definecolor{BLACK}{gray}{0}
 \definecolor{WHITE}{gray}{1}
 \definecolor{RED}{rgb}{1,0,0}
 \definecolor{GREEN}{rgb}{0,1,0}
 \definecolor{BLUE}{rgb}{0,0,1}
 \definecolor{CYAN}{cmyk}{1,0,0,0}
 \definecolor{MAGENTA}{cmyk}{0,1,0,0}
 \definecolor{YELLOW}{cmyk}{0,0,1,0}
}
\usepackage{dcolumn}% needed for some tables
\usepackage{bm}% for math
\makeatother
\begin{document}
\newcommand{\norm}[1]{\left\lVert#1\right\rVert}

\title{Gradient-based optimal control of open quantum systems using quantum trajectories and automatic differentiation}

\author{Mohamed Abdelhafez}
\email{abdelhafez@uchicago.edu}
\affiliation{The James Franck Institute and Department of Physics, University of Chicago, Chicago, Illinois 60637, USA}

\author{David I. Schuster}

\affiliation{The James Franck Institute and Department of Physics, University of Chicago, Chicago, Illinois 60637, USA}
\author{Jens Koch}
\affiliation{Department of Physics and Astronomy, Northwestern University, Evanston, Illinois 60208, USA}

\date{\today }

\begin{abstract}

We present a gradient-based optimal-control technique for open quantum systems that utilizes quantum trajectories to simulate the quantum dynamics during optimization. Using trajectories allows for optimizing open systems with less computational cost than the regular density matrix approaches in most realistic optimization problems. We introduce an improved-sampling algorithm which minimizes the number of trajectories needed per optimization iteration. Together with employing stochastic gradient descent techniques, this reduces the complexity of optimizing many realistic open quantum systems to the complexity encountered with closed systems. Our optimizer harnesses automatic differentiation to provide flexibility in optimization and to suit the different constraints and diverse parameter regimes of real-life experiments. We utilize the optimizer in a variety of applications to demonstrate how the use of quantum trajectories significantly reduces the computation complexity while achieving a multitude of simultaneous  optimization targets. Demonstrated targets include high state-transfer fidelities despite dissipation, faster gate times and maximization of qubit-readout fidelity while maintaining the quantum non-demolition nature of the measurement and allowing for subsequent fast resonator reset.

\end{abstract}
\maketitle
\section{Introduction}
Controlling the time evolution of quantum systems to achieve certain optimization targets has been a crucial task in a host of contexts, including quantum chemistry \cite{Khaneja_2005,Maday_Turinici_2003,Reich_Ndong_Koch_2012,Zhu_Botina_Rabitz_1998}, NMR spectroscopy \cite{de_Fouquieres_Schirmer_Glaser_Kuprov_2011,Herb_2005,Khaneja_Brockett_Glaser_2000,Nielsen_Kehlet_Glaser_Khaneja_1996,Tosner_Vosegaard_Kehlet_Khaneja_Glaser_Nielsen_2009}, molecular physics \cite{Palao_Kosloff_2002,Shapiro_Brumer_2003} and quantum information theory \cite{Borzi_Salomon_Volkwein_2008}. In addition, many applications of optimal control have been emerging in the field of quantum computing in order to realize quantum gates \cite{Blume-Kohout_Caves_Deutsch_2002, Schirmer_2009,Chow_DiCarlo_Gambetta_Motzoi_Frunzio_Girvin_Schoelkopf_2010} of different qubit realizations including superconducting qubits \cite{kelly2014optimal,Sporl_Schulte-Herbruggen_Glaser_Bergholm_Storcz_Ferber_Wilhelm_2007,Fisher_Helmer_Glaser_Marquardt_Herbruggen_2010,Egger_Wilhelm_2013} and ion traps \cite{Singer_Poschinger_Murphy_Ivanov_Ziesel_Calarco_Kaler_2010,Timoney_Elman_Glaser_Weiss_Johanning_Neuhauser_Wunderlich_2008,Zhao_Babikov_2008}. There exist different theoretical approaches to implementing optimal control in closed quantum systems, but the main commonly used algorithms are Gradient Ascent Pulse Engineering (GRAPE) \cite{Khaneja_Reiss_Kehlet_Schulte-Herbruggen_Glaser_2005}, as well as Krotov \cite{krotov1995global,Reich_Ndong_Koch_2012} and other monotonically converging gradient algorithms \cite{Ohtsuki_Turinici_Rabitz_2004,Maday_Turinici_2003,gollub2008monotonic}.

Studying open quantum systems where dissipation and dephasing effects are included requires the simulation of open dynamics. Typically, this can be described by a Markovian Lindblad master equation \cite{BRE02}. A key difference between open and closed dynamics is that the open case propagates the quantum state as a density matrix instead of a state vector. Most of the working closed-system algorithms can be generalized to the open case but would have to propagate density matrices of dimension $d^2$ instead of state vectors of dimension $d$. Open algorithms include open versions of GRAPE \cite{Khaneja_Reiss_Kehlet_Schulte-Herbruggen_Glaser_2005} and Krotov methods \cite{krotov1995global}. Density matrix centered algorithms were successfully used in some applications of small system size \cite{Gordon_Kurizki_Lidar_2008,Tai_Lin_Goan_2014,Chou_Huang_Goan_2015,Jirari_2009}. While traditional propagation requires consistent matrix multiplications and exponentials of superoperators of dimension $d^2 \times d^2$, some more recent approaches rely on propagating the density matrix through different integration methods \cite{Gutknecht_2007,Goerz_2015,boutin2017resonator}. Notwithstanding the achieved improvements, these techniques are still based on storing and propagating matrices of size at least $d \times d$. In addition, propagation is generally achieved  through complex and often computationally tedious integration techniques. Most of the applications demonstrated with these techniques were limited to moderate Hilbert space sizes.

A useful alternative for simulating the dynamics of open quantum systems is to employ quantum trajectories. Quantum trajectories describe the effect of the environment on the system by a stochastic Schr{\" o}dinger equation (SSE). This SSE governs how the evolution of the system is conditioned on the measurement processes of the environment \cite{BRE02,Wiseman_Milburn_2009}. Every trajectory carries information about the dynamics and the average over many trajectories reproduces the master equation solution. Trajectories offer a promising computationally efficient approach to open optimal control since generating every trajectory requires only sparse matrix-vector propagation. Recently, the use of trajectories has been proposed in Krotov-based optimization \cite{Goerz_Jacobs_2018} for a specific choice of optimization target. However, there has been little work on the use of quantum trajectories in gradient-based optimal control, since the stochastic nature of trajectories makes it difficult to provide analytical forms of gradients for gradient-based algorithms. Here, we present a gradient-based trajectories optimization technique that significantly enhances the simulation complexity and is flexible with respect to the choice of optimization targets. The foundation of this technique relies on automatic differentiation. 

In our previous work \cite{Leung_Abdelhafez_Koch_Schuster_2017}, we invoked the similarities between quantum optimal control and deep neural networks \cite{Haykin_1994,Hecht-Nielsen_1989}. Both  rely on linear algebra processing and gradient descent techniques to minimize a certain error function. Therefore, we may utilize the automatic differentiation concept \cite{Baydin_Pearlmutter_Radul_Siskind_2015,Bartholomew-Biggs_Brown_Christianson_Dixon_2000,Rall_1981} that is commonly used in neural network machine learning applications to build our trajectory-based optimizer. The main advantage is that analytical forms of trajectory gradients for the various  optimization constraints are not needed to be hard-coded in the algorithm. An additional advantage is that a reduced number of trajectories could be used per optimization iteration, since  loading the gradients of a subset of the total data still leads to convergence. This process is known as stochastic gradient descent (SGD), and is intensively used in machine learning applications \cite{Ketkar_2017,Bottou_2010,Wang_Yang_Min_Chakradhar_2017,Zinkevich_Weimer_Li_Smola_2010}.

\section{Theory}
\subsection{Quantum Trajectories}
\label{traj}
Whenever a quantum system interacts with its environment, it is subject to dissipation and dephasing processes, and must be treated as an open system. In general, the Hamiltonian of such an open system interacting with its environment is given by
\begin{equation}
H_{\text{tot}} = H + H_{\text{env}} + H_{\text{int}},
\end{equation}
where $H$ is the Hamiltonian of the system of interest, $H_{\text{env}}$ the Hamiltonian of the environment and $H_{\text{int}} $ the corresponding interaction Hamiltonian. Several generic assumptions about the nature of the environment interaction lead to the Lindblad master equation \cite{BRE02} which is the evolution equation for the reduced density matrix of the system, $\rho = \Tr_{\text{env}} (\rho_{\text{tot}})$:
\begin{equation}
\frac{\partial \rho}{\partial t} = - i [H,\rho] - \frac{1}{2} \sum_l \gamma_l \big[ c_l^{\dagger} c_l \rho + \rho c_l^{\dagger} c_l - 2 c_l \rho c_l^{\dagger} \big] = -i \mathcal{L} \rho.
\end{equation}
Here, $\{c_l\}$ is a set of so-called jump operators, describing relaxation and dephasing emerging from the interaction with the environment, with associated rates $\{ \gamma_l \}$. $\mathcal{L}$ denotes the Liouville superoperator and we have set $\hbar=1$. In the following, we will assume that the initial state in the evolution is a pure state of the system, $\rho(0) = \ket{\psi_0}\bra{\psi_0}$.

Following the evolution of open quantum systems requires propagation of the full $d \times d$  density matrix, where $d$ is the system Hilbert space dimension. Numerically, this may be realized by expressing the density matrix as a vector and propagating it via matrix exponentials of the $d^2\times d^2$ superoperator. As Hilbert space size increases, this process can require heavy computational resources in comparison to the closed-system dynamics governed by propagators of size $d \times d$.

An alternative approach to simulating the dynamics of open quantum systems consists of quantum trajectories: a stochastic-sampling approach which simulates $m$ independent trajectories. Each trajectory undergoes dynamics of a complexity similar to that of a closed system, and thus only requires propagators of size $d \times d$ which helps reduce the computational cost. There are different equivalent unravelings of the master equation into a set of trajectories \cite{Wiseman_Milburn_2009}. Here, we utilize the simplest of them: the unraveling into quantum-jump trajectories. The generation of these trajectories is summarized as follows \cite{daley2014quantum}:
\begin{enumerate}
\item Discretize the total evolution time into small time steps $dt$.
\item Propagate the initial state $\ket{\psi_0}$ with the effective non-Hermitian Hamiltonian 
\begin{equation}\label{eq:heff}
H_{\text{eff}} = H - \frac{i}{2} \sum_l \gamma_l \ c_l^{\dagger} c_l.
\end{equation} 
The norm $\norm{\ket{\psi(t)}}$ of the resulting state will decay over time.
\item Generate a uniformly-distributed random number $r\in [0,1)$.
\item Keep propagating with $H_\text{eff}$ until $\norm{\ket{\psi(t)}}^2$ reaches $r$. Once reached, randomly select one jump operator from $ \{ c_l \}$, according to probabilities $\propto \bra{\psi(t)} \gamma_l \, c_l^{\dagger} c_l \ket{\psi(t)}$. Apply the jump operator to the state, and normalize the result.
\item Repeat steps 3-4.
\end{enumerate}
The expectation value of any Hermitian operator $A$ can be obtained by averaging over a sufficient number $M$ of trajectories, \begin{equation}
\langle A \rangle (t) = \Tr [A \rho(t)] \approx \frac{1}{M} \sum_m \bra{\psi_m(t)} A \ket{\psi_m(t)}
\end{equation}
with statistical error $\sigma_A \propto \frac{1}{\sqrt{M}}$ \cite{Goerz_Jacobs_2018}. 

Using quantum trajectories to simulate the dynamics of open systems has several advantages. First, the complexity of computations is $\mathcal{O} (M d^3)$ instead of $\mathcal{O} (d^6)$ as would be needed to propagate the full density matrix. [Note: matrix multiplication of matrices of size $d \times d$ is $\mathcal{O}(d^3)$.] In addition, we can use the sparsity of matrices that typically arise in most quantum applications to significantly lower the complexity of matrix-vector multiplication to $\mathcal{O}(d^2)$. Therefore, for large Hilbert space dimensions where $M < d^3$ (with $d \propto 2^n$ growing exponentially with n, the number of qubits), the use of quantum trajectories can significantly improve the complexity. While $M$ must be large enough to overcome the statistical noise in the trajectories outcomes, we will propose schemes to reach statistical convergence with a lower number of trajectories. In many practical cases quantum jumps turn out to be rare. We propose a protocol that takes advantage of the rarity of jumps, and only requires computation of a smaller set of trajectories, hence reducing the complexity even further. In addition, quantum trajectories reduce the memory requirements for calculating the forward evolution, as superoperators are never stored in memory explicitly; only propagators of size $d \times d$ are calculated instead. Finally, since trajectories are independent of each other, this method is also highly parallelizable. Different trajectories can be run on different computational nodes in parallel, thus reducing overall computation time.
%So, even if m trajectories are needed for good statistics, if enough resources exist, all trajectories could be simulated in parallel which makes the whole simulation need just the equivalent time for one closed system evolution. This is not possible to do in the standard way of propagating the density matrix where if the system size increases, one can't utilize different nodes to help the computation run faster or run at all. If the computational resources of one node are not enough for the density matrix propagation, there is no way to combine the resources of another node to help, so quantum trajectories offer an alternative to not being able to run the simulation at all by dividing it into smaller tasks that multi-nodes can tackle. \\ 

To present how quantum trajectories can be used in optimal control, we first review the relevant gradient-based optimal-control techniques. 

\subsection{Open GRAPE}
In standard optimal control theory, time evolution is discretized into $N$ time steps of  duration $dt$.  During each time step $j$, the Hamiltonian consists of adjustable control parameters $u_{kj}$, assumed to be constants for each period $dt$. These control parameters multiply a set of control Hamiltonians $H_k$ which are added to the constant system Hamiltonian (the so-called drift Hamiltonian) $H_0$,
\begin{equation}
H_j \equiv H(j\,dt) = H_0 + \sum_k u_{kj} H_k.
\end{equation}
%In that case, \textcolor{black}{[And what case is that? The flow of text is a bit confusing to me here. The paragraph sounds like it starts with closed GRAPE, but then suddenly switches to the open case.]} the corresponding Liouville superoperators $\mathcal{L}_j$ can be separated again into drift (including dissipation) and control superoperators as \begin{equation}
%\mathcal{L}_j = \mathcal{L}_0 + \sum_k u_{kj} \mathcal{L}_k 
%\end{equation} 
This set of discretized Hamiltonians is used to propagate the initial state(s) at $t = 0$ towards the final state(s) at $t = t_N$. In the case of open quantum dynamics, the standard way to perform such propagation is through Liouville superoperators $\mathcal{L}_j$ which can be separated again into drift (including dissipation) and control superoperators as \begin{equation}
\mathcal{L}_j = \mathcal{L}_0 + \sum_k u_{kj} \mathcal{L}_k. 
\end{equation} 

The goal is to determine the optimal control parameters $u_{kj}$ that minimize a certain cost function $C$. Gradient descent algorithms can be utilized as long as gradients of $C$ with respect to the control parameters $u_{kj}$ can be calculated. For instance, an iterative approach may be used in which the control parameters are updated via
\begin{equation}
u_{kj} \rightarrow u_{kj} -  \epsilon \frac{\partial C}{\partial u_{kj}},
\end{equation}
where $\epsilon$ is the update step size. Several algorithms exist to change $\epsilon$ adaptively at every iteration for better convergence. More complex gradient descent methods such as ADAM \cite{Kingma_Ba_2015} may also be utilized.

A key cost function in open quantum optimal control theory is the distance between the target density matrix $\rho_T$ and the propagated density matrix $\rho_N$ at the final $N$th time step. The corresponding infidelity cost function can be expressed as
\begin{equation}
C = 1- \Tr[\rho_T \rho_N].
\end{equation}
The open GRAPE method \cite{boutin2017resonator} relies on calculating the analytical gradients $\frac{\partial C}{\partial u_{kj}}$. For that purpose, consider the final density matrix $\rho(T)$, obtained after $N$-fold application of the propagation superoperators 
\begin{equation}
\rho_N = \Lambda_N \Lambda_{N-1} \dots \Lambda_1 \rho_0
\end{equation}
where the propagator at time step $j$ is
\begin{equation}
\Lambda_{j} = \text{exp}(- i \mathcal{L}_j \  dt ).
\end{equation}
The main approximation in the standard calculation of  gradients in open GRAPE comes from expressing the derivative of every $\Lambda_j$ to the first order in $dt$ as 
\begin{equation} \label{open_approx}
\frac{\partial \Lambda_j}{\partial u_{kj}}   \approx - i\,  dt\, \Lambda_j \mathcal{L}_k. 
\end{equation}
which ignores higher order terms that include commutators of $\Lambda_j$ and $\mathcal{L}_k$.
While this first-order approximation is standard in the current GRAPE literature when using a small $dt$, higher-order expansions with better accuracy have been considered \cite{de_Fouquieres_Schirmer_Glaser_Kuprov_2011}.
\begin{comment}
as
\begin{equation}
\begin{split}
\frac{\partial \Lambda_j}{\partial u_{kj}} = \Lambda_j \big[ -i dt \mathcal{L}_k  + \frac{(dt)^2}{2!} [ \Lambda_j, \mathcal{L}_k]   \\ + \frac{i (dt)^3}{3!} [ \Lambda_j, [ \Lambda_j, \mathcal{L}_k ] ] + \dots \big]
\end{split}
\end{equation}
\end{comment}

Working to first order in $dt$, the analytical gradient in this case can be calculated to be
\begin{equation}
\label{grape}
\frac{\partial C}{\partial u_{kj}}  \approx i \ dt\, \Tr ( \lambda_j [H_k,\rho_j] )
\end{equation}
where $\rho_j = \rho(j\,dt)=\Lambda_j \Lambda_{j-1} \dots \Lambda_1 \rho(0)$ is the initial density matrix propagated to time step $j$ and $\lambda_j = \Lambda_{j+1}^{\dagger} \Lambda_{j+2}^{\dagger} \dots \Lambda_{N}^{\dagger} \rho_T$ is the target density matrix backward-propagated to the same time step $j$.

Therefore, utilizing GRAPE to optimize open quantum systems necessitates the calculation of $\rho_j$ and $\lambda_j$ at every time step $j$. This is realized by propagating both the initial and target density matrices using matrix exponentials of superoperators of dimension $d^2 \times d^2$. Hence, the usual GRAPE optimization process has all the computational and memory drawbacks discussed in section \ref{traj}. Therefore, using quantum trajectories is a potential area of improvement for open GRAPE when the Hilbert space dimension is large.
However, extending gradient-based algorithms to deal with quantum trajectories is not entirely straightforward due to the randomness inherent in every time step of each trajectory, as we shall discuss next. % makes it tricky to obtain the analytical gradient during the dynamic evolution. 

\subsection{Gradients in Quantum Trajectories}
Consider the problem of efficiently calculating analytical gradients for quantum trajectories by comparing with the case of closed GRAPE. Each quantum trajectory consists of the time evolution of a pure state which is described by a stochastic Schr\"odinger equation \cite{Wiseman_Milburn_2009}. In this aspect, it resembles the closed-system evolution under the ordinary Schr\"odinger equation. In closed systems, variations of the GRAPE algorithm \cite{Khaneja_Reiss_Kehlet_Schulte-Herbruggen_Glaser_2005} describe how to obtain gradients analytically. A key strength of these algorithms is that they enable the calculation of analytical gradients in an efficient way during the simulation of the system evolution.

To illustrate this, consider the following simple state-transfer cost function in the case of a closed system, 
\begin{equation} \label{eq:infid}
C = 1 - |\langle\psi_T|\psi_{N}\rangle|^2,
\end{equation}
where $\ket{\psi_N} = U_N U_{N-1} \dots U_1 \ket{\psi_0}$ is the final system state, $\ket{\psi_T}$ is the corresponding target state and $U_j = \exp(-i H_j dt)$ is the unitary propagator at time step $j$. 
Similar to the open-GRAPE case, the closed-GRAPE gradients can be calculated to first order in $dt$, resulting in \cite{Khaneja_Reiss_Kehlet_Schulte-Herbruggen_Glaser_2005}
\begin{equation}
\label{closed_grad}
\frac{\partial C}{\partial u_{kj}} = -2 \ dt \iim \bigg [ 
\langle \psi_{T_{j}} | H_k  | \psi_{j}\rangle  \langle \psi_T | \psi_{N}\rangle^* 
\bigg ]
\end{equation}
where $\ket {\psi_{T_{j}}} =  \prod_{j'=j+1}^{N} U_{j'}^{\dagger} \ket{\psi_T}$ is the target state back-propagated in time to time step $j$,  and $\prod$ is understood to produce a time-ordered product. Eq.\ \eqref{closed_grad} indicates that an efficient gradient calculation method which does not need caching of intermediate states or propagators can be realized as follows: 
\begin{enumerate}
\item Complete forward evolution given the current values of the control parameters $u_{kj}$ to obtain $\ket{\psi_N}$.
\item Back-propagate both $\ket{\psi_N}$ and $\ket{\psi_T}$ one time step backwards using $U_{N}^{\dagger}$ and calculate the gradients at this time step using Eq. \eqref{closed_grad}.
\item Repeat step 2 to back propagate to all time steps using $U_{N-1}^{\dagger}, U_{N-1}^{\dagger} , \dots, U_{1}^{\dagger}$ and calculate the corresponding gradients for all $j$.
\end{enumerate}

Turning back to the case of quantum trajectories, we note two main differences distinguishing quantum trajectories from the evolution of closed quantum systems. First, the propagation in every time step is no longer unitary due to the non-Hermitian part of $H_{\text{eff}}$, see Eq.\ \eqref{eq:heff}. Second, quantum trajectories involve randomness due to the intermittent occurrence of quantum jumps. The propagation,
\begin{equation}\label{eq:psin}
%\begin{split}
\ket{\psi_N} = \frac {M_N M_{N-1} \cdots M_1 \ket{\psi(0)}} {\norm{M_N M_{N-1} \cdots M_1 \ket{\psi(0)}}}
%\end{split}
\end{equation}
is described using the non-unitary propagator
\begin{equation}\label{Ms1}
M_j = 
\exp(\textstyle -i H_j dt - \frac{dt}{2} \sum_l \gamma_l c_l^{\dagger} c_l )
\end{equation}
in the absence of a jump, or
\begin{equation}\label{Ms2}
M_j = c_l,
\end{equation}
if a jump occurs in decoherence channel $l$.
Imitation of the cache-free gradient calculation via back-propagation of states will generally fail for quantum trajectories. Back-propagation cannot be  achieved by $M_j^\dag$ anymore due to non-unitarity. Even worse, most realistic jump operators do not even have an inverse as they represent irreversible dynamical changes of the system, e.g., the decay of an excitation. This leads to the necessity of caching intermediate information during the numerical simulation and takes away an important advantage of using fully analytical forms of GRAPE.

Besides the need for caching, obtaining an analytically closed form for gradients is considerably more tedious for  quantum trajectories than in the case of closed unitary evolution. In part, this is due to the need for explicit normalization of propagated states, see Eq.\ \eqref{eq:psin}. Different from closed evolution, the final single-trajectory state $\ket{\psi_N}$ now depends on the control parameters not only via the propagators $M_j$ but also via the normalization factor $F_N = \norm{M_N M_{N-1} \cdots M_1 \ket{\psi(0)}}$ in the denominator of Eq.\ \eqref{eq:psin}. As a result, gradients with respect to the control parameters thus require both $\partial M_j/\partial u_{kj}$ and $\partial F_N/\partial u_{kj}$. This makes it generally much more cumbersome to obtain analytical gradients and use them in an efficient way. For the simple infidelity cost function in Eq. \eqref{eq:infid}, we provide a detailed discussion of the analytical gradients in Appendix \ref{append:analytical}.

 Therefore, one way to better handle the cumbersome matter of gradient calculation for quantum trajectories given any desired cost function, is to use automatic differentiation instead of relying on analytical gradient forms which, if existent, do not generally support efficient implementation without caching anyways.

\subsection{Automatic Differentiation}
Automatic differentiation is a central concept in machine learning, utilized for the general optimization of cost functions \cite{Baydin_Pearlmutter_Radul_Siskind_2015,Bartholomew-Biggs_Brown_Christianson_Dixon_2000,Rall_1981}. As we have demonstrated in previous work on closed-system optimization \cite{Leung_Abdelhafez_Koch_Schuster_2017}, its framework can also be utilized in quantum optimal control problems. The main idea of automatic differentiation is to systematically apply the chain rule of differentiation to relate a given cost function to the control parameters and calculate the corresponding gradients automatically. The process of expressing the optimization problem in the automatic-differentiator language creates a computational graph that is used to track the gradients in a backward fashion. For the reader unfamiliar with these concepts, details and examples of the use of automatic differentiation are provided in Appendix \ref{append:ad}. 
In the following, we discuss how automatic differentiation can be implemented in the case of optimization based on quantum trajectories. 

\section{Implementation}
\subsection{Conditional Graphing using TensorFlow}
\begin{figure} [htbp]
  \centering \includegraphics[width=1\linewidth]{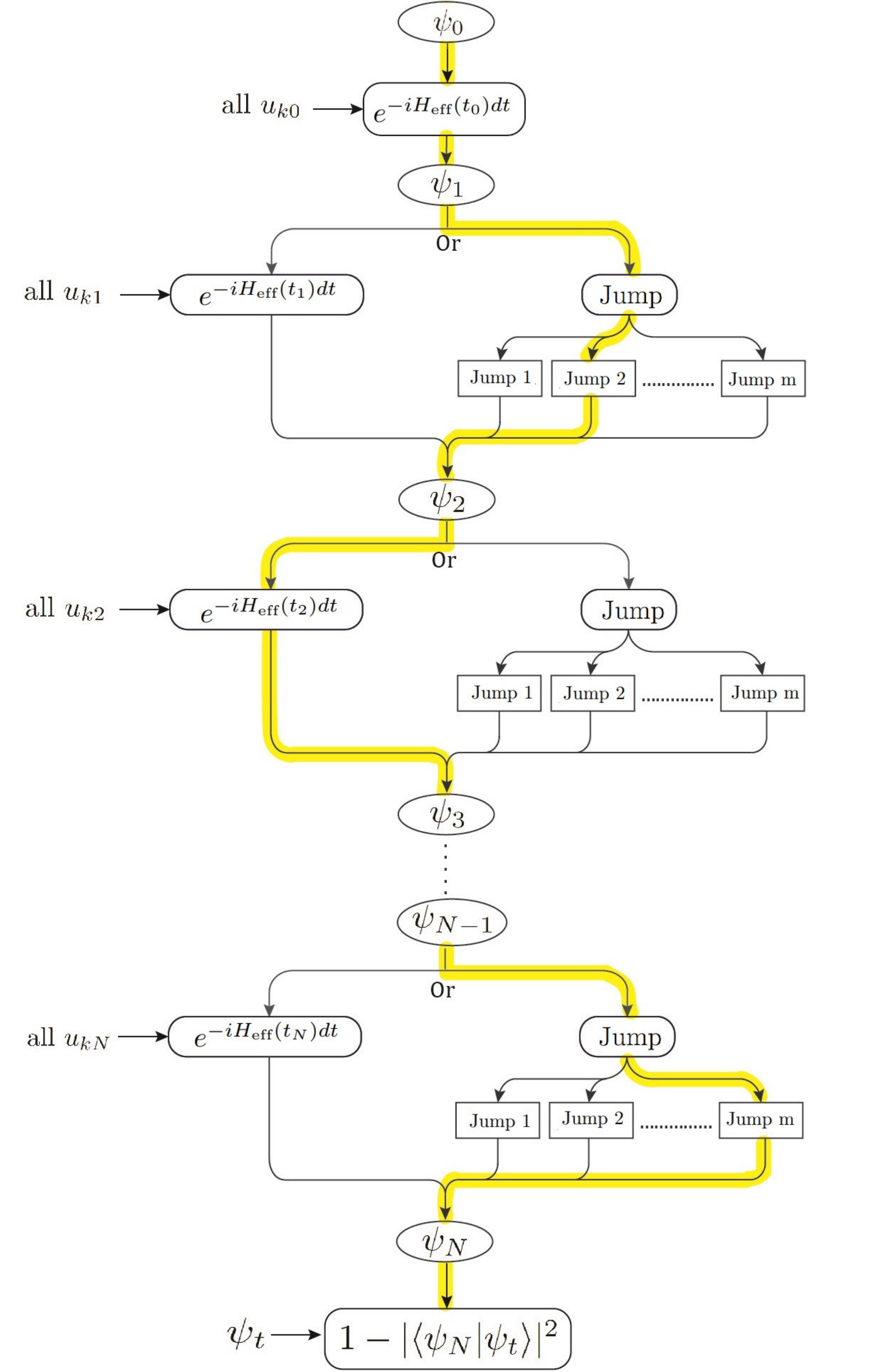}
  \caption{The computational graph of a quantum trajectory evolution must be conditional and allow for all possibilities of the forward path. Each forward path is only determined at runtime through the choice of random numbers entering the trajectory generation. At every timestep, the evolution either proceeds via the non-unitary Hamiltonian $H_{\text{eff}}$ or through a jump $\in$ $\{c_l\}$ from one of the possible $m$ jump channels. The yellow path is an example of the many possible trajectories.% that is chosen at runtime to calculate gradients for.
  \label{figure:cond_graph}}
\end{figure}

In order to use automatic differentiation for quantum trajectories, we must define the computational graph of every trajectory in terms of basic matrix operations defined by the differentiator. One challenge in this process is the randomness inherent in each quantum trajectory which prevents the computational graph from having a fixed structure. Instead, the relationship between the inputs and the cost function is only determined at run time by the choice of the random numbers entering the individual trajectory. Hence, our differentiator is equipped with the flexibility of dynamically creating the computational graph of each trajectory -- a scenario called conditional graphing. Fig.\ \ref{figure:cond_graph} illustrates the type of conditional graph needed for quantum trajectories.

We implement our open quantum optimizer using the TensorFlow library for automatic differentiation and machine learning \cite{Abadi_Others_2016}. Developed by Google's machine intelligence research group, it allows for conditional graphing and dynamically setting the size of the computational graph at runtime. TensorFlow is easily integrated within Python and has a comprehensive basic set of operations for matrix algebra with predefined gradients. It also supports the use of GPUs in a straight-forward manner and allows one to select which computations to perform on available CPUs vs.\ GPUs. This enables better use of computational resources to balance speed and memory usage of the simulation \cite{Leung_Abdelhafez_Koch_Schuster_2017}. In addition, TensorFlow has support for distributed learning, by running different parts of the computational graph or multiple copies of the same graph on different machines in parallel. This is crucial for the implementation of quantum trajectories, as we will discuss below.

\subsection{Custom Cost Functions}
So far, we have primarily focused our discussion on the infidelity cost function defined in Eq.\ \eqref{eq:infid}. However, one key advantage of automatic differentiation is the ability to define a broad variety of cost functions without the need to calculate analytical gradients manually. Our implementation takes advantage of that to allow the user to specify a weighted sum of a multitude of cost functions according to the needs of optimization. Cost functions need not be functions of only the final evolved state $\psi_N$. Instead, they may depend on all intermediate states $\{\psi_j\}$, or could impose constraints on the control parameters themselves.

\begin{table}[htbp]
\caption{Examples of single trajectory cost functions implemented in our optimizer} % title of Table
\centering % used for centering table
\begin{ruledtabular}
\begin{tabular}{c p{0.43\linewidth} } % centered columns (4 columns)
 %inserts double horizontal lines
 Cost Function&  \centering Explanation  \tabularnewline % inserts table
\hline %heading
 % inserts single horizontal line
$C_1 = 1 - |\langle\psi_T|\psi_{N}\rangle|^2$ &  Infidelity between final state $\psi_{N}$ and target state $\psi_T$ \\ [1ex]
% [1ex] adds vertical space
 %inserts single line
$C_2 = \displaystyle\sum_j |\langle\psi_f|\psi_{j}\rangle|^2$ &  Intermediate occupation of forbidden state $\psi_f$\\ [1ex]
$ C_3 = \displaystyle\sum_j \langle\psi_j|O|\psi_{j}\rangle$ &  Integrated expectation value of operator $O$ \\ [1ex] 
$C_4 = \displaystyle\sum_{k, j} | u_{kj} - u_{k j-1} |^2$ &  First derivatives of the control parameters\\ [1ex]
$C_5 = $ $\displaystyle\sum_{k, j} | u_{kj+1} -2u_{kj} + u_{k j-1} |^2$ &  Second derivatives of the control parameters   \\ [1ex]
$C_6 = \displaystyle\sum_{k,j} |u_{kj}|^2$ & Control pulse power \\ [1ex]
$C_7 = \displaystyle\sum_{k,j}|(1 - e^{- \frac{(j - \frac{N-1}{2})^2}{2\sigma^2}}) u_{kj}|^2 $ & Deviation from a Gaussian envelope
\end{tabular}
\label{table:cost_fns} % is used to refer this table in the text
\end{ruledtabular}
\end{table}

Important examples of relevant cost functions we implement are shown in Table \ref{table:cost_fns}. For the cost functions $C_1, C_2$ and $C_3$, the optimizer monitors and optimizes over intermediate and/or final physical quantities, like the intermediate/final occupation of some state or the intermediate/final expectation value of a physical operator. By contrast, cost functions $C_4, C_5, C_6$ and $C_7$ ensure that the resulting control pulses are smooth and experimentally realizable. The total cost function in any optimization instance is an appropriate linear combination of these individual contributions, $C = \sum \alpha_i C_i$ where the weight  coefficients $\alpha_i$ are determined empirically depending on the desired relative strengths of the different constraints.

\subsection{Techniques for Handling Trajectories}
Another very important consideration for the optimizer is the ability to implement the number of trajectories needed for the simulation efficiently, which differs according to the size and nature of each optimization problem. Fortunately, it is not always necessary to exhaustively sample the full statistics in each single iteration of the optimization process. Instead, it may be sufficient to generate a smaller number of trajectories which only partially represent the statistics. In the case where only a single trajectory is used per iteration, the procedure is known as Stochastic Gradient Descent (SGD) \cite{Ketkar_2017,Bottou_2010,Wang_Yang_Min_Chakradhar_2017,Zinkevich_Weimer_Li_Smola_2010}. This generally leads to non-monotonic convergence, often requires more iterations and can produce noisy pulses. A better way for our model is to include a moderate number of trajectories in each iteration so that every iteration is based on a batch of data. This procedure, where gradients applied in each iteration are generated by a batch of trajectories, is known as mini-batch SGD \cite{shalev2011stochastic}. %In all cases, the optimizer needs to allow for efficient navigation of the needed number of trajectories according to the needs of the optimization.

Using quantum trajectories allows for flexibility in the ways simulations could be implemented since trajectories could be constructed independently of each other. Whether to generate trajectories in series and/or in paralel, and how many trajectories should be grouped together, are examples of questions that the optimizer should address on a case by case basis, depending on the specific optimization problem. Hence, we implement different ways of grouping and propagating trajectories in our optimizer to better match the nature of each optimization problem. In particular, the following three techniques are used to handle trajectories generation.

\subsubsection{Improved-Sampling Algorithm}
We present an algorithm for generating a balanced sample of trajectories per iteration using only a fraction of the needed number of trajectories. This leads to improving the complexity of the calculation while keeping convergence smooth and time efficient. 

Suppose $m_{\text{tot}}$ is the batch size, i.e., the number of trajectories to be used in every iteration within mini-batch SGD. SGD leads to convergence even with relatively small batch sizes without significantly reducing the convergence speed as will be shown in the applications. Therefore, SGD allows for limiting $m_{\text{tot}}$ to a small number which helps further reducing the complexity.

There are two potential limitations to this approach. First, control pulses will typically be noisy if a very small number of trajectories is used for every gradient update. To prevent this from happening, we utilize cost functions $C_4 - C_7$ to ensure that stochastic noise in the pulses is canceled. Second, there is an increase in the number of iterations required to reach convergence. To circumvent this limitation, we introduce a technique which only implements a subset of size $m_{\text{sim}}$ of the intended $m_{\text{tot}}$ trajectories. From this subset, we generate a better balanced sample for every iteration that makes convergence smooth and fast and also hugely reduces the complexity of the problem, especially if jumps are rare. 

 In many practical cases,  dissipation and dephasing time scales are much longer than the time scale governing the dynamics of the system. For example, many optimized gates on superconducting qubits may take no more than $0.01 - 0.1 \mu\text{s}$, while decoherence times of the qubit are orders of magnitude larger. In that case, dissipative terms in the Liouvillian will have a smaller impact on the dynamics and quantum jumps are less likely to  occur. As a result, inside a statistically representative batch of trajectories, many trajectories will be identical to the no-jump trajectory. To eliminate the redundancy of generating the no-jump trajectory many times, our implementation allows for performing a test run that first generates the no-jump trajectory. Then, using the fact that the state norm of a quantum trajectory monotonically decreases over time \cite{BRE02} in the absence of jumps, we extract the final norm of the no-jump trajectory and identify it with the no-jump probability for the total evolution time. From then on, we only generate trajectories which do include jumps by controlling the range of the random numbers $r$. Consequently, all remaining generated trajectories in a batch are representatives of jump trajectories and the total number of trajectories to be generated can be reduced: if jumps are sufficiently rare, $m_{\text{sim}}$ is only a small percentage of  $m_{\text{tot}}$, with the precise fraction given by the probability of jumps. Finally, we perform a weighted average of the gradients of both the no-jump trajectory and the generated jump trajectories according to the calculated jump/no-jump probability. This algorithm is summarized below:
\begin{enumerate}
\item Generate the no-jump trajectory (random number $r = 0$). 
\item Save the no-jump gradients $g_{\text{nj}}$ from this trajectory. 
\item Extract the norm of the final state of this no-jump trajectory, $p = \langle\psi_N|\psi_{N}\rangle$ and assign it as the no-jump probability.
\item If $m_{\text{tot}}$ trajectories are needed, generate only \begin{equation} m_\text{j} = \lceil (1-p) m_\text{tot}\rceil \end{equation} jump trajectories with $r \in [p,1)$, where $\lceil x \rceil$ denotes the integer ceiling of $x$. The reduced range for $r$ guarantees at least one jump will happen since the norm will definitely drop below $r$. After the jump, $r$ is reset to the full range $[0,1)$ to simulate potential additional jumps. 
%Hence $m_{\text{sim}} = m_j +1$ for every initial starting state.
%(If $m_j$ is small (or zero), we could increase it to generate more jump statistics that will then be rebalanced by the no jump ones.)
\item Calculate the averaged jump gradient $g_{\text{j}}$ from the $m_\text{j}$ trajectories.
\item Calculate the net gradient of all $m_{\text{tot}}$ trajectories,
\begin{equation}
g = (1-p) g_{\text{j}} + p\, g_{\text{nj}}
\end{equation}
\end{enumerate}

If jumps are rare, this algorithm saves significant resources by only generating $m_{\text{sim}} = m_\text{j}+1$ trajectories instead of $m_{\text{tot}}$. In addition, instead of just implementing randomly chosen trajectories every iteration, the algorithm will always generate a balanced sample every iteration/batch that represents jumps and no jumps with correct weights. This leads to a much smoother and faster convergence than the case where $m_{\text{tot}}$ is a collection of (generally unbalanced) random events.

As a concrete example, consider the case of a problem with jump probability $(1-p)=10\%$, and with a small batch size $m_{\text{tot}} = 10$ to reduce computational costs. Then, conventionally all 10 trajectories used in a given iteration could be identical to the no-jump trajectory. Since the occurrence of jumps is not uniform from iteration to iteration, convergence would be non-monotonic, and properly sampling the dynamics would require a bigger number of iterations. However, using the algorithm described above, each iteration  is enforced to contain the same balanced statistics between jumps and no jumps, even for small sample size.  Therefore, the improved-sampling algorithm is a crucial element of our optimizer that renders the complexity of an open-system problem similar to that of a closed system, if jumps are rare, as will be demonstrated in the applications.

\subsubsection{Matrix-Vector Exponential and Clustering Trajectories}
Another important computational bottleneck is the evaluation of the matrix exponential required for state propagation via the effective Hamiltonian. Our implementation eliminates the need to calculate the matrix exponential through matrix-matrix multiplication. Instead, when propagating a state $V_j =\ket{\psi_j}$ to $V_{j+1} = \ket{\psi_{j+1}}$ through the matrix exponential $e^A$, where $A= -i (H_{\text{eff}})_{j+1} dt$, we use an iterative Taylor series to reexpress the propagation as 
\begin{align}
\label{eq:mat-vec}
V_{j+1} &= e^A V_j = (1 + A + \frac{A^2}{2!} + \frac{A^3}{3!} + \cdots) V_j \\\nonumber
&= V_j + A V_j + \frac{1}{2!} A (A V_j) + \frac{1}{3!} A (A(A V_j)) + \cdots
\end{align}
The new state can thus be calculated by an iterative series of matrix-vector multiplications, % without ever having to construct the matrix exponential itself. 
reducing the complexity of propagation at every time step from $\mathcal{O}(n_{\text{Taylor}} d^3)$ which is required by matrix-matrix multiplication to $\mathcal{O}(n_{\text{Taylor}} d^2)$, where $n_{\text{Taylor}}$ is the number of Taylor expansion terms kept in the simulation. 

Using the improved-sampling algorithm, a batch of $m_{\text{sim}}$ initial states needs to be propagated per iteration. Instead of sequential generation of each trajectory, storing all gradients and then  averaging them in the end, the above matrix-vector implementation allows for simultaneous propagation of a number of vectors. The procedure described in equation (\ref{eq:mat-vec}) can be generalized from $V_j$ representing a $d \times 1$ vector, to denoting a matrix of size $d \times m_{\text{sim}}$. Using this idea, we can cluster trajectories together and process them in a faster way than running them in series \cite{liniov2017increasing}. Clustering trajectories is particularly useful if the needed number of trajectories is much smaller than the Hilbert space dimension for big-sized problems or if the Hilbert space dimension is relatively small and there are a lot of unused resources when propagating one trajectory at a time. In our implementation, we treat  the desired number of trajectories to be combined in each iteration (the batch size) as an adjustable parameter. It may be specified at runtime, and enables dynamically selecting the size of the computational graph. It also allows for combining the effects of several batches and then applying their averaged gradients to the control parameters. Those two features together give reasonable flexibility in dividing the needed number of trajectories into batches of combined trajectories and in matching  available computational resources to the concrete size and nature of each optimization problem.

\subsubsection{Parallelization of Trajectories}
While generating the needed statistics from a smaller number of clustered trajectories saves memory and runtime, a prime advantage of quantum trajectories is their high degree of parallelizability. We can utilize parallelization to further improve the efficiency and flexibility of the optimizer. Our implementation uses TensorFlow's distributed learning features to run a number of different clustered trajectories on different nodes in parallel. We have built an interface between the SLURM manager  for operating clusters \cite{yoo2003slurm} and TensorFlow allowing for the use of both synchronous and asynchronous training. Here, synchronous training refers  to the situation when all compute nodes must finish their mini-batches together and their gradients are then averaged. Asynchronous training, by contrast, gives every compute node the ability to update the control parameters once its batch gradients are ready. Both types of training are of practical importance, depending on the needed statistics per iteration for stable convergence. Our implementation uses ``between graph replication" which means that every node builds its own identical version of the computational graph. Then, communication between graphs is coordinated through a chief worker machine.

\section{Showcase Applications}
\subsection{Transmon Qubit State Transfer }
 To illustrate the improved-sampling algorithm, we consider a transmon qubit with $n = 4$ levels, initialized in the ground state $\ket{g}$. We wish to transfer the system to the first excited state $\ket{e}$. For the transmon, we assume a frequency difference between ground and excited levels of $\omega_{ge}/2\pi = 3.9$ GHz and an anharmonicity of $\alpha /2 \pi $ =  - 225 MHz.  The control Hamiltonians $\{ H_x, H_z \}$ couple to the $x$ and $z$ degrees of freedom of the qubit, so that the net Hamiltonian is given by
\begin{equation}
 H = \omega_{ge} b^{\dagger} b + \frac{\alpha}{2} b^{\dagger} b (b^{\dagger} b -1) + \Omega_x (t) (b^{\dagger}+ b) + \Omega_z (t) b^{\dagger} b.
\end{equation} 
Here, $b$ and $b^{\dagger}$ are ladder operators for the transmon excitation level, truncated at an appropriate level ($n = 4$ in our case). The qubit is coupled to a heat-bath environment, resulting in relaxational dynamics with characteristic time $T_1$. Working at zero temperature, the corresponding jump operator is $b$. We utilize the following cost functions: $C_1$ to maximize the fidelity between the final evolved state and $\ket{e}$, $C_4 - C_7$ to generate smooth realizable control pulses and $C_2$ to forbid the occupation of the $n$-th level so that the truncation of the transmon levels remains valid. 

\begin{figure}
  \centering \includegraphics[width=0.95\columnwidth]{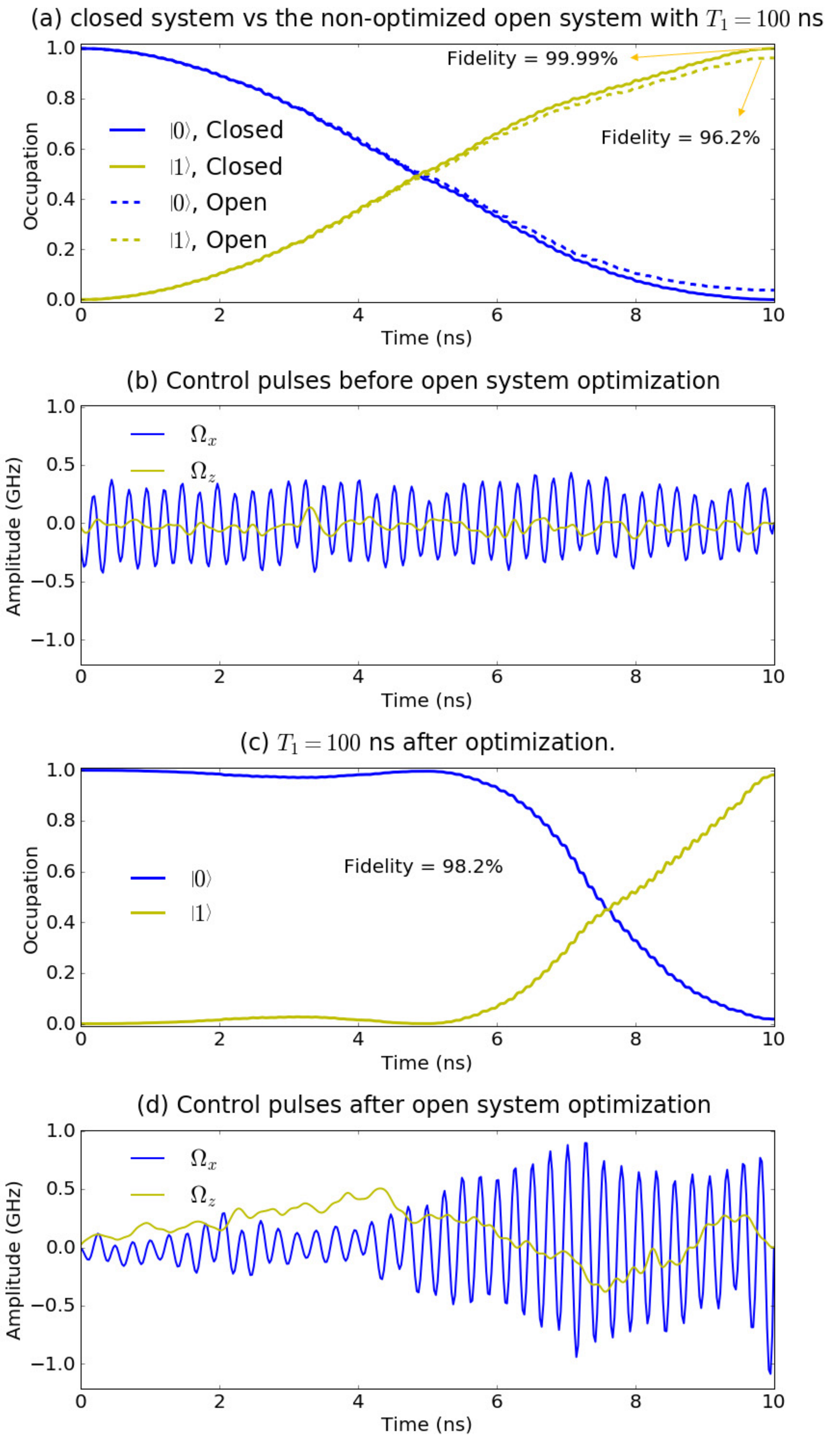}
  \caption{ Optimizing state transfer from the ground to the first excited level in a lossy transmon qubit. Panels (a) and (c) show the occupation of both levels monitored over the pulse period while panels (b) and (d) show the pulse trains obtained from optimization. (a) In the absence of relaxation, the optimized solution reaches a closed-system fidelity of $99.99\%$. Then when this solution is applied in the presence of relaxation processes ($T_1= 100\,\text{ns}$), the resulting state-transfer fidelity drops to $96.2\%$. (b) The corresponding pulse sequences. (c) Results from trajectory-based GRAPE. The new optimized solution raises the state-transfer fidelity to $98.2\%$ (d) The optimized pulse minimizes relaxation effects by delaying the pulse as much as possible, then rapidly performing the transfer using increased power.
  \label{figure:T1}}
\end{figure}

First, we study the effect of relaxation on the results from the state-transfer optimization. In the closed-system case where  relaxation is absent ($T_1 \rightarrow \infty$), we readily achieve state-transfer fidelities of 99.99$\%$ within a total evolution time of $T = 10$\,ns, see  Fig.\ \ref{figure:T1}. However, if the qubit is fairly lossy (taking, for example, $T_1 = 100$\,ns), the previously determined pulse train only achieves a state-transfer fidelity of 96.2$\%$ since occupation of the $\ket{e}$ level is inevitably subject to dissipation. Re-running the optimization in the presence of $T_1$ processes, we succeed in increasing the fidelity to 98.2$\%$, so we gain around $2\%$ even for this example of a rather lossy qubit, see Fig.\ \ref{figure:T1}.

Inspection of the results reveals that the optimizer aims to minimize the detrimental effects of relaxation by delaying the state-transfer operation as much as possible. This way, the total time period over which the state $\ket{e}$ is occupied, is reduced and there is, hence, a smaller time window where the system is sensitive to decay. While the closed-system case utilizes the whole time to perform the state transfer, the open dynamics state transfer is much more asymmetric in time, reflecting the asymmetry between the ground and the first excited state as far as decoherence is concerned. Note that the optimization cannot fully bring the fidelity back to $99.99 \%$ because the occupation of $\ket{e}$ is limited by the relaxation $ e^{-t/T_1}$. Note that this result is also useful in optimizing the final time for the state transfer, since the obtained pulse indicates the total time required for the state transfer, which is even relevant for the choice of the total time in the closed-system analysis.  

\begin{figure}
  \centering \includegraphics[width=\columnwidth]{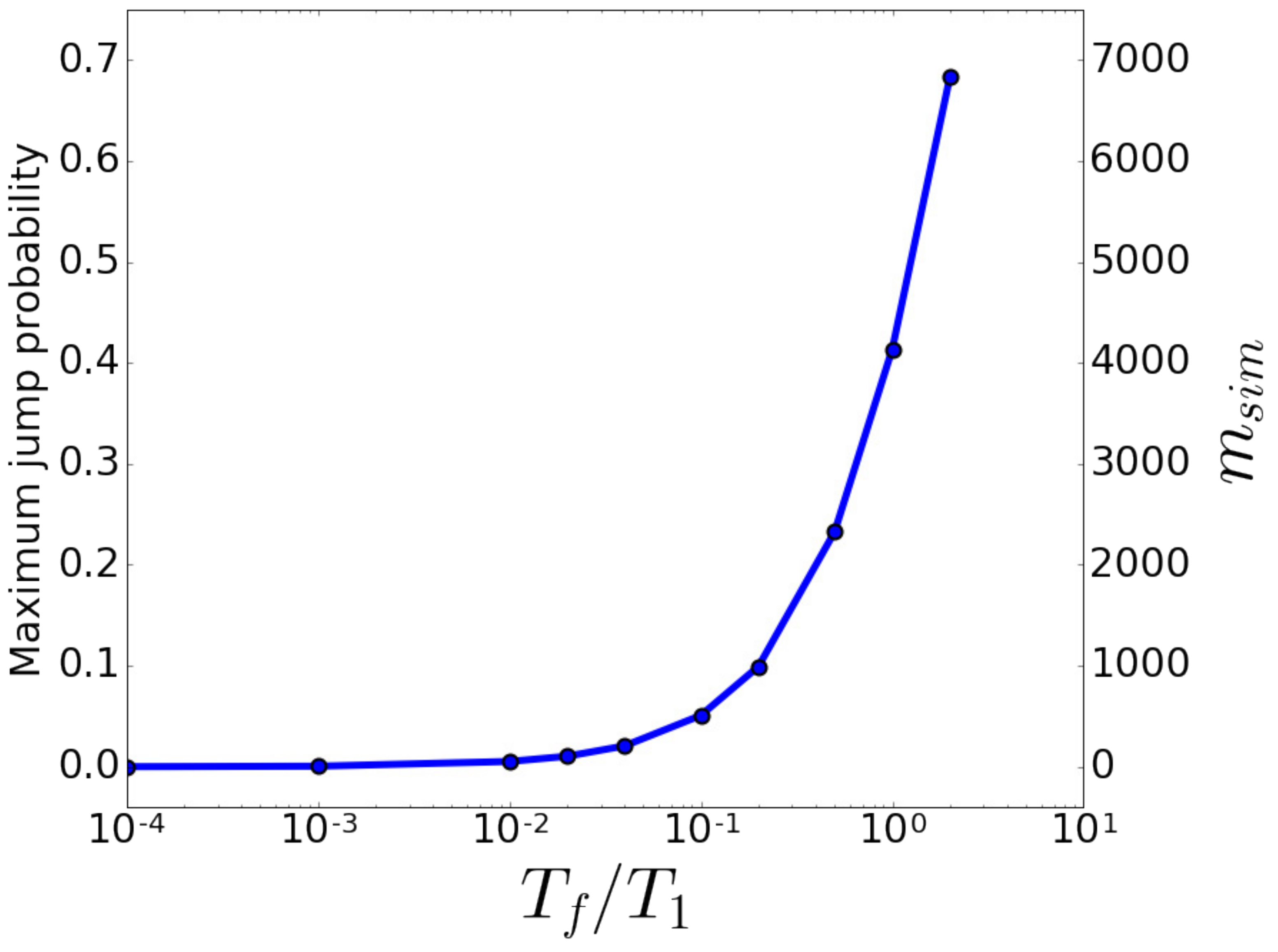}
  \caption{Maximum jump probability for the driven transmon qubit as a function of the simulation time $T_f$ measured in units of the relaxation time $T_1$. This represents the fraction of the number of trajectories that the improved-sampling algorithm will generate every iteration. In most realistic cases, the fraction does not exceed $1-5\%$, allowing for a significant reduction in computational costs.
  \label{figure:js}}
\end{figure}

Next, we discuss the choices of $m_{\text{tot}}$ and $m_{\text{sim}}$ needed for the optimization. 
Deploying our improved-sampling algorithm, we can perform the trajectory-based optimization at nearly the same computational cost as in the closed-system case. We ran simulations using different values of $T_1$ to obtain the probability of jumps for each case. Note that this probability changes with the iterations of the optimization since the Hamiltonian changes according to the updated control pulses. So, focusing on the maximum value we get for the probability of jumps, we get the results in Fig.\ \ref{figure:js} for different values of the ratio $T_f/T_1$ where $T_f$ is the final time of the state transfer.

\begin{figure} 
  \centering \includegraphics[width=\columnwidth]{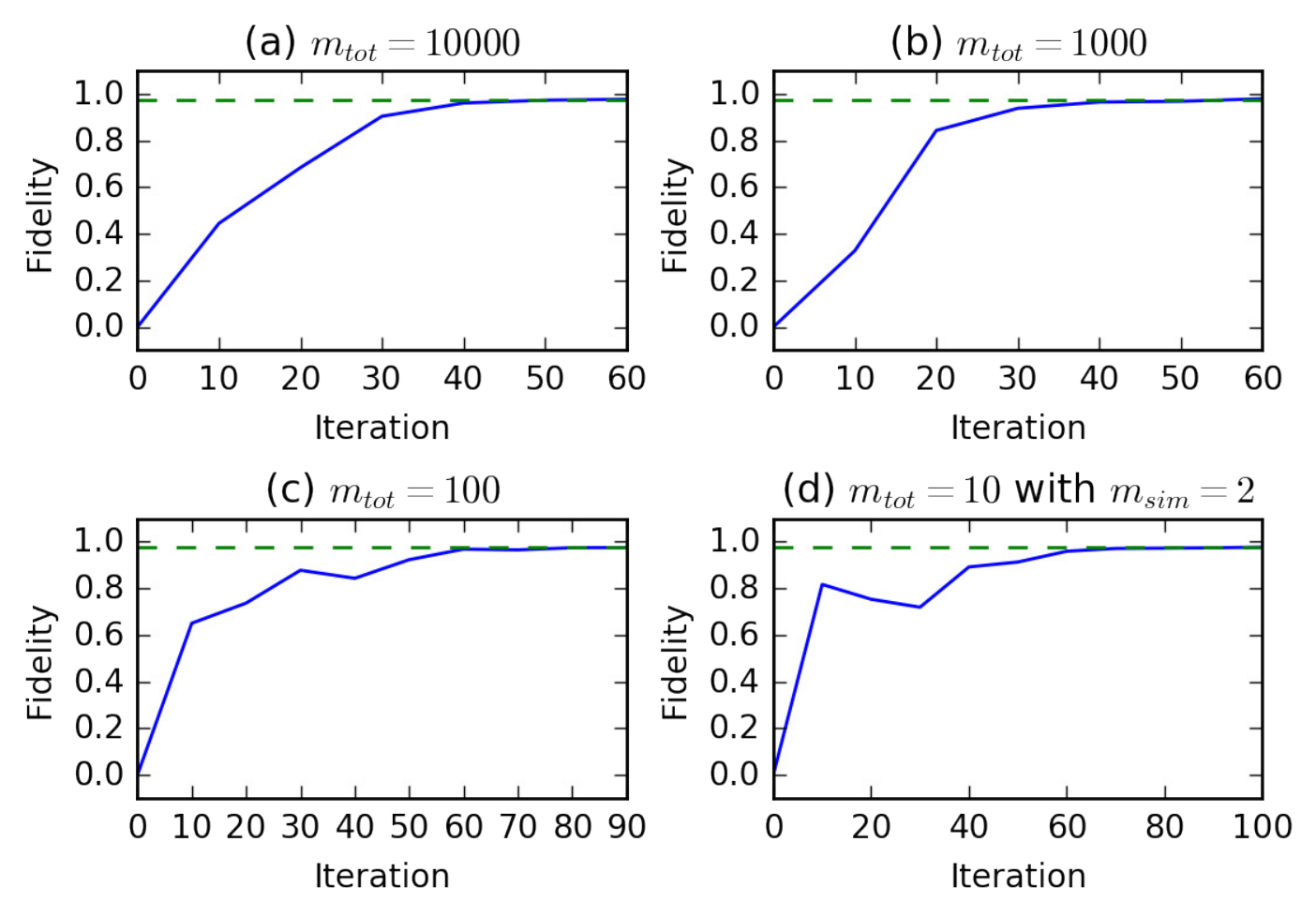}
  \caption{Convergence of the optimization algorithm to a target fidelity of $97.5\%$ for different values of the batch size $m_{\text{tot}}$, using the improved-sampling algorithm. 
  \label{figure:mtot}}
\end{figure}

As Fig.\ \ref{figure:js} shows, even if the needed number of trajectories $m_{\text{tot}}$ is 10,000, only a small number of trajectories is actually simulated. For example, for $T_f/T_1 < 10^{-2}$, the probability of jumps is less than 0.6$\%$, allowing us to obtain the desired sampling by generating merely 60 trajectories. Even if the final time is increased to $0.1\,T_1$ like in Fig.\ \ref{figure:T1}, the probability of jumps is around 5$\%$ which is still a small fraction. 
In most of realistic applications, the sample size $m_{\text{tot}}$ does not have to be as large as 10,000, since using our algorithm allows for good convergence even for much smaller total numbers of  trajectories. We compare the convergence for different choices of $m_{\text{tot}}$ using our algorithm in Fig.\ \ref{figure:mtot}. 

As shown in Fig.\ \ref{figure:mtot}(d), we obtain good convergence even for a small batch size of $m_{\text{tot}} = 10$ and $m_{\text{sim}} = 2 $ (i.e., only generating two trajectories per iteration) within around 100 iterations, reaching a target fidelity of 97.5$\%$. By comparison, for a significantly larger batch size of $m_{\text{tot}} = 10,000$, we reach the same fidelity within 60 iterations. Therefore, we only need to double the number of iterations to simulate the system using a thousand times fewer trajectories per iteration. In addition, every iteration will be  much faster since it only includes running two trajectories. 

Running the same optimization problem for batch size $m_{\text{tot}} = 10$ but without the improved-sampling algorithm leads to convergence to the same target fidelity after 320 iterations. Therefore, using the improved-sampling algorithm needs only two trajectories per iteration for a total number of 200 trajectories to reach convergence, in comparison to 3,200 trajectories needed when not employing the algorithm.

\subsection{Lambda System Population Transfer}
\subsubsection{Problem Overview}
\begin{figure} [hptb]
  \centering \includegraphics[width=0.85\columnwidth]{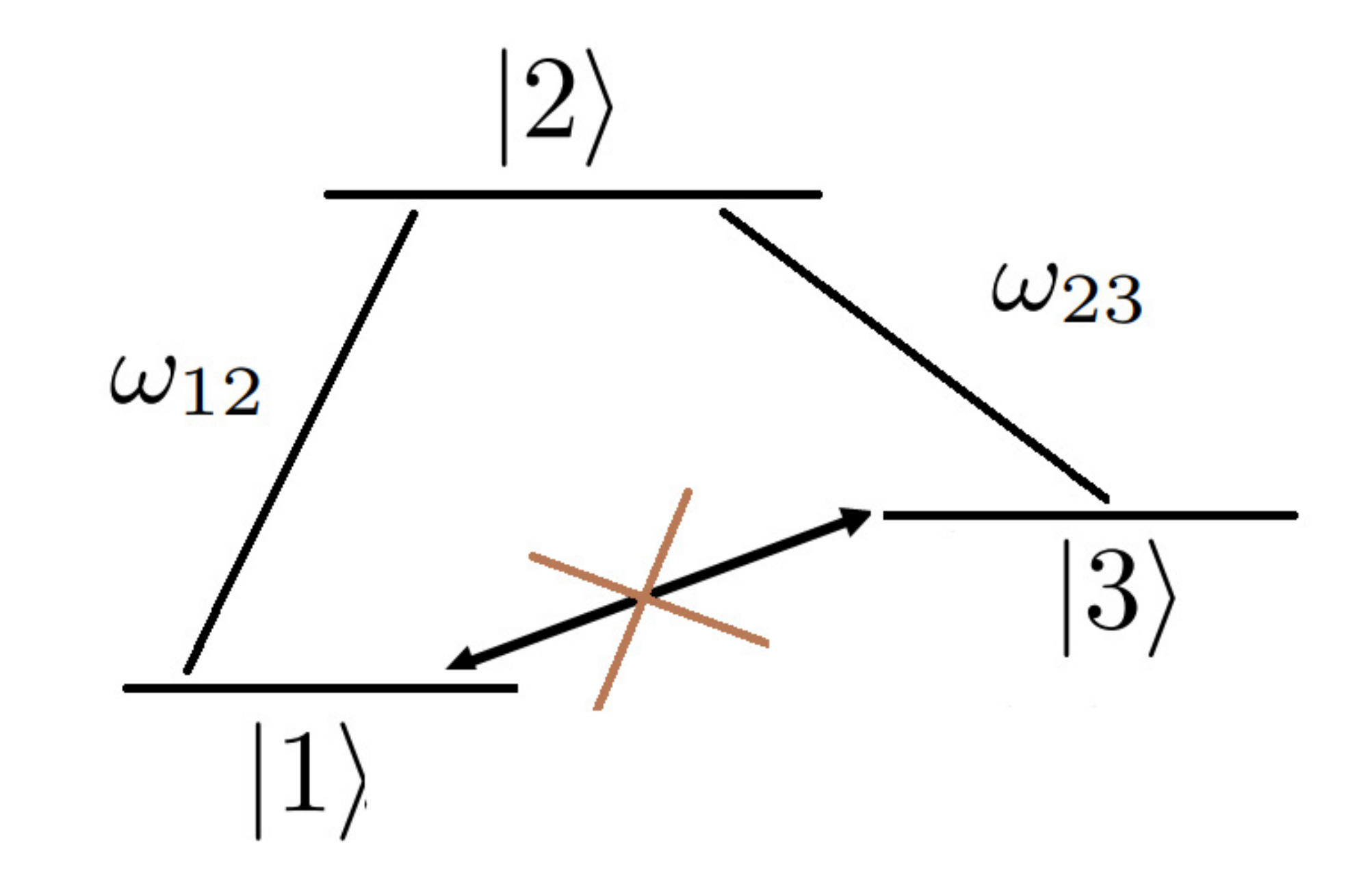}
  \caption{The $\Lambda$ system consists of ground state $\ket{1}$ and meta-stable excited state $\ket{3}$, as well as an intermediate lossy state $\ket{2}$. The frequencies $\omega_{12}$ and $\omega_{23}$ are distinct and no direct matrix element exists between $\ket{1}$ and $\ket{3}$. Hence, state transfer between them must invoke the intermediate state $\ket{2}$. The system is driven with a pulse $\zeta(t)$ that is optimized to maximize the state-transfer fidelity. 
  \label{figure:Lambda}}
\end{figure}

Having shown how the improved-sampling algorithm helps reduce the problem complexity for the toy example of a transmon state transfer, we next consider a more realistic case of driving nearly forbidden transitions in a three-level $\Lambda$ system as shown in Fig. \ref{figure:Lambda}. 

Two levels of it, $\ket{1}$ and $\ket{3}$, are stable, i.e., the direct transition between $\ket{1}$ and $\ket{3}$ is forbidden. The third, intermediate level $\ket{2}$ can decay to either of the former two states and direct matrix elements allow one to drive the $\ket{1} \longleftrightarrow \ket{2}$ and $\ket{3} \longleftrightarrow \ket{2}$ transitions. The goal is to transfer the system from one stable state to the other while avoiding significant occupation of the intermediate state which is subject to errors from spontaneous dissipation. This application arises in many realistic quantum systems, for example in circuit quantum electrodynamics (cQED). Inducing transitions in such a $\Lambda$ system is crucial for protected qubits like the heavy-fluxonium \cite{earnest2018realization} and the 0-$\pi$ qubit \cite{brooks2013protected}.

\subsubsection{Protocol 1: Raman Transitions}
We will compare two existing protocols to induce the desired transition. The first is the two-photon Raman transition in which the system is driven by two off-resonant pulses of amplitudes $\Omega_1$, $\Omega_2$, and detunings $\delta_1 = \delta_2 = \delta$. For convenience, it is assumed that each pulse couples solely to a single transition between the intermediate state and one of the stable states. Adiabatic elimination is an approximation that may be performed if the detuning $\delta$  is considered large \cite{brion2007adiabatic}. This approximation assumes small occupation of the intermediate level, and hence, the system may be treated as an effective two-level system. Under this assumption, this system is shown to be effectively driven by a Rabi oscillation of an effective frequency that is proportional to both $\Omega_1$ and $\Omega_2$. In this limit, the dissipation of the intermediate level is negligible since the level remains largely unoccupied during the transfer.

This approximation, however, is only valid if $\Delta = \delta_1 +\delta_2 \gg \Omega_1, \Omega_2 $. If the transfer is desired to occur over shorter time scales, then the required effective Rabi frequency, and hence $\Omega_1$ and $\Omega_2$, must be increased. This may invalidate the adiabatic elimination condition, since needed larger detunings will cause the frequencies used by the two pulses to be too far from the transitions frequencies. In that case, it is not guaranteed that each pulse drives a single transition separately as assumed by the adiabatic elimination. Hence, this will result in the system deviating from the simplified picture of an effective two-photon process as we will show in the results subsection below. 

\subsubsection{Protocol 2:  STIRAP}
The second commonly used protocol is the stimulated-Raman-adiabatic-passage (STIRAP) method. This involves the application of two partially overlapping pulses which adiabatically keep the $\Lambda$ system in a superposition of the two stable states without occupying the intermediate level \cite{gaubatz1990population}. First, a Stokes pulse [with amplitude $\Omega_S$(t)] is used to couple the two unoccupied states $\ket{2}$ and $\ket{3}$. Then, a pump pulse [with amplitude $\Omega_P$(t)] couples states $\ket{1}$ and $\ket{2}$ in such a way that makes direct transition from $\ket{1}$ to $\ket{3}$ possible. STIRAP assumes that the rotating wave approximation (RWA) is valid which causes the system to have three-time dependent eigenstates, one of them has no projection in the $\ket{2}$ state \cite{vitanov2017stimulated} at all times. This eigenstate is labeled the dark state $\ket{\psi_d}(t)$. If the pulses are changed adiabatically such that $\Omega_S$(t) is smoothly turned off while $\Omega_P$(t) peaks and then turns off, the system will remain in $\ket{\psi_d(t)}$ and reaches the desired state at the end without occupying the intermediate state at all as ensured by remaining in the dark state.

The required adiabaticity of the transfer, however, necessitates strong limits on the time needed for STIRAP transfer. In particular, the minimum time where pulses overlap is inversely proportional to the peak pulse amplitude used (in frequency units)\cite{bergmann1998coherent}. The constant of proportionality is estimated from experimental data, and for around 95$\%$ STIRAP efficiency, it is estimated to be 10 \cite{vitanov2017stimulated}. This poses a restriction on how fast the transfer can be accomplished. While the peak pulse amplitude could be increased to achieve shorter times, high pulse powers would eventually violate the RWA and result again in population of the intermediate level. 
Therefore, to achieve fast transfer between the two stable states in a $\Lambda$ system, existing techniques will inevitably involve partial population of the intermediate state which endangers the transfer fidelity because of its dissipative nature. 

Recently, enhanced protocols which use shortcuts to adiabaticity \cite{zhou2017accelerated} were presented to improve the speed of STIRAP while maintaining high transfer fidelities. Our optimal control results are comparable to those techniques in terms of speed and transfer efficiency.

\subsubsection{Simulation Details}
We will utilize the full Hamiltonian of the system (without applying the RWA) in order to allow for more optimization freedom. Different from the Raman and STIRAP protocols described above, we do not restrict the pulses to a single frequency component. Instead, we work with a single pulse of general form that can couple to both transitions. The Hamiltonian then takes the form
\begin{equation}
H = \sum_{i=1}^3  \omega_i \ket{i}\bra{i} + \zeta(t) (H_{12} + \alpha H_{23})
\end{equation}
with the definition
\begin{equation}
H_{ij} = \ket{i}\bra{j} + \ket{j}\bra{i}.
\end{equation}
Here, $\alpha$ is a factor that relates the matrix elements of the two transitions. For our simulations, we used the sample values $\omega_1/2\pi = 0$ GHz, $\omega_2/2\pi = 5$ GHz, $\omega_3/2\pi = 1.8$ GHz and $\alpha = 1$. We associate state $\ket{2}$ with a relatively short relaxation time $T_1 = 20$\,ns due to decay into either of the two stable states, and we set the target transfer time to 10\,ns, while  limiting the amplitude of $\zeta(t)$ to an experimentally reasonable maximum value of $3/2\pi$\,GHz. 

Note that with these parameters, the STIRAP protocol requires at least around $21$\,ns of the Stokes and pump pulses to overlap, which is not even the full time of the protocol.  Hence, the optimizer will search for a solution that is at least twice as fast as STIRAP, but still maintains low occupation of $\ket{2}$ and high transfer fidelity.

\subsubsection{Results}
\begin{figure}
  \centering \includegraphics[width=\columnwidth]{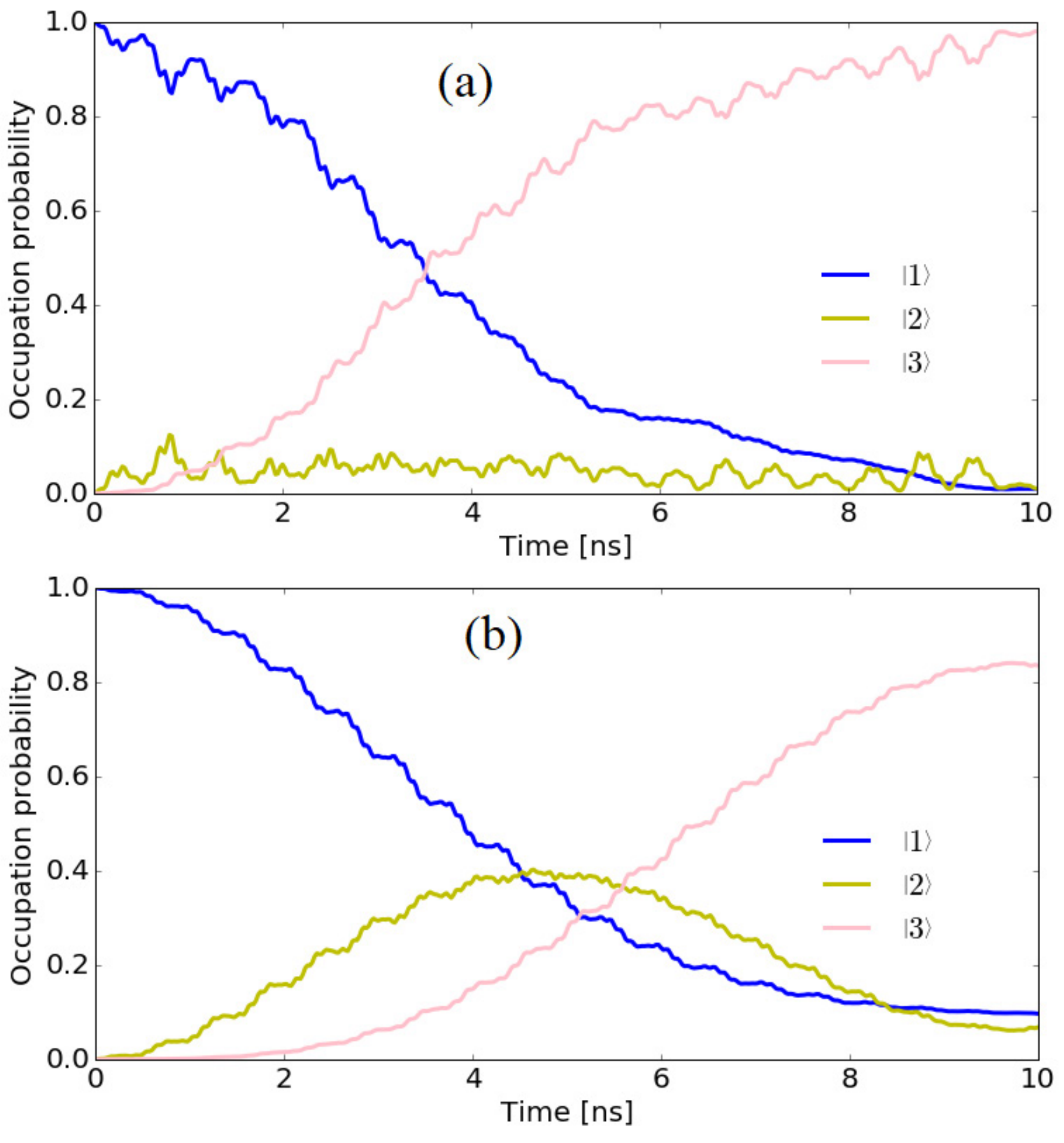}
  \caption{ Results of driving the Lambda system with the optimized pulses vs. the usual two-photon Raman pulses. (a) The optimized pulse achieves a transfer fidelity of 98.0$\%$ by using several tones to properly limit the occupation of the lossy $\ket{2}$ state.  (b) The two-photon Raman transition fails to limit the occupation of $\ket{2}$, and hence can only reach a fidelity of 84.1$\%$ within the parameter regime of our simulation.
  \label{figure:lambda_results}}
\end{figure}

With a maximum jump probability of around 10$\%$ (as calculated during simulation), the improved-sampling algorithm allows for a 90$\%$ reduction of the number of trajectories generated. % and hence reduced complexities were achieved like the previous application. 
Around 10 trajectories were used per iteration, thus accounting for an effective number of simulated trajectories of 100 per iteration. Convergence was reached with an optimized fidelity of $98.0\%$. To compare the solution against the two-photon Raman transition using the same total transfer time and pulse amplitude, trials with different values for the two pulse detunings and amplitudes were performed.  The best solution yields only a fidelity of $84.1\%$ and shows significant occupation of the intermediate level. The results are summarized in Fig. \ref{figure:lambda_results}. In summary, our optimizer succeeds in high-fidelity population transfer in a $\Lambda$ system with a comparatively short transfer time for which standard protocols are significantly less efficient.

\subsection{Quantum Non Demolition (QND) Readout of a Transmon Qubit Allowing Fast Resonator Reset}
\label{sec:readout}
The previous two applications showed how our optimizer works in scenarios where jumps are rare. In the following application, we will deal with a situation where jumps are not rare and focus on the capabilities of the optimizer in a relatively big Hilbert space where a density-matrix--based approach would be difficult to implement.

\subsubsection{Problem Overview}
In cQED, a common technique to measure the state of a transmon qubit is to couple it to a readout resonator. This is described by the generalized Jaynes-Cummings model Hamiltonian
\begin{equation}
H = \omega_r a^{\dagger} a + \omega_q b^{\dagger} b  + \frac{1}{2} \alpha b^{\dagger} b (b^{\dagger} b -1)+ g (a^{\dagger} b + a b^{\dagger}) + H_\zeta(t),
\label{eq:Hfull}
\end{equation}
where $\omega_r$ and $\omega_q$ are the bare resonator and qubit frequencies, respectively, and $a$ is the lowering operator for photons inside the resonator. $b$ and $b^{\dagger}$ are the ladder operators for the transmon excitation number, truncated at an appropriate level. There are two jump operators for the total system, $a$ for photon loss with a rate of $\kappa$ and $b$  qubit decay with rate $\gamma = \frac{1}{T_1}$. $H_\zeta(t)$ is the resonator drive term responsible for generating the readout, and takes the form
\begin{equation}
H_\zeta(t) = \zeta(t) (a e^{i \omega_d t}+ a^{\dagger}e^{-i \omega_d t}) 
\end{equation}
where $\omega_d$ is the drive frequency.

For the purpose of qubit readout, the circuit parameters are chosen such that the system is in the dispersive regime ($\Delta = |\omega_q - \omega_r| \gg g$) \cite{wallraff2004strong,blais2004cavity}. In that case, the effective frequency of the resonator is AC-Stark shifted by a value that is dependent on the qubit state. Hence, by driving the resonator at either of the two shifted frequencies, the amplitude-response of the readout tone can distinguish between the 0 and 1 states of the qubit. Alternatively, by driving at the bare resonator frequency, the qubit state can extracted from information carried in the phase of the readout \cite{blais2004cavity}.

\subsubsection{1st Optimization Target: High Readout Fidelity}
An appropriate metric for successful readout is the single-shot readout fidelity $\mathcal{F}$ \cite{magesan2015machine}
\begin{equation}
\label{eq:fid}
\mathcal{F} = 1- \frac{p(0|1) + p(1|0)}{2}
\end{equation}
where $p(a|b)$ is the probability of measuring the qubit in state $a$ given it was prepared in state $b$. There are two main factors that limit the readout fidelity $\mathcal{F}$: noise in the measurement and qubit-decay processes that can make an excited-state trajectory look very similar to a ground-state one. While noisy readout could be mitigated by increasing the measurement time and hence enhancing the ability to average out the noise, this worsens the probability for spurious qubit decay during the measurement, leading to a decrease of the readout fidelity.

Existing readout protocols involve multiplying the readout signal by a filter function, and integrating it over the measurement time. The integration result is then compared against a set threshold in order to identify it with one of the underlying qubit states 0 or 1. Several filters exist to maximize $\mathcal{F}$ including the optimal linear filter \cite{gambetta2007protocols} which we will utilize here.
 
 One key area of improvement that we pursue here is the choice of the pulse $\zeta(t)$ that maximizes the fidelity. In most experiments, $\zeta(t)$ is taken to have a square-pulse envelope, rendering the time-dependent drive sinusoidal with a constant amplitude that is varied in order to maximize $\mathcal{F}$. Using our optimizer, we open up the possibility of a wider variety of pulse trains including multiple frequency components which may yield higher fidelities while maintaining readout speed.
 (Further details of the implementation of trajectories and calculation of the fidelity are explained in appendix \ref{append:diffusive}).
 
 \subsubsection{2nd Optimization Target: Overcoming Dressed Dephasing and Obtaining Fast Resonator Reset}
 \textcolor{black}{Increasing the readout-pulse power can yield higher fidelities as it counteracts experimental noise. However, excessively high powers lead to increased photon occupations throughout the measurement which has several drawbacks.}

 \textcolor{black}{One drawback pertains to the ability to perform the measurement in a manner that facilitates a fast cavity-reset process afterwards,  allowing for new measurements to be started with minimum downtime. There are several protocols for both passive and active reset of the cavity \cite{mcclure2016rapid,bultink2016active}. The time it takes to reset the cavity depends on how many photons are left in the cavity by the end of measurement. Ref.\ \cite{boutin2017resonator} shows that there is a power law relating the speed limit of resonator reset and the number of leftover resonator photons. For example, doubling the number of photons in the resonator requires an increase in the active resonator-reset time of at least $57\%$. Therefore, not limiting the cavity occupation number can significantly slow down subsequent resonator reset.}
  
 \textcolor{black}{In addition, high photon numbers cause another significant  problem, namely dressed dephasing \cite{boissonneault2008nonlinear} which results from higher-order corrections to the dispersive approximation, usually ignored in the regime of small photon numbers. As the number of photons increases, the quantum non-demolition (QND) nature of the readout is jeopardized by these extra dephasing channels.}

\textcolor{black}{In line with these insights, we utilize the flexibility of setting targets in our optimizer to search for a readout pulse that maintains high levels of fidelity whilst keeping the resonator occupation as low as possible.}
 
 \subsubsection{3rd Optimization Target: QND Measurement}
 \textcolor{black}{One crucial element of the readout that needs to be maintained is its quantum non-demolition (QND) behavior. Typical readout measurements within the dispersive regime are expected to keep the qubit state unchanged for sufficiently low photon occupation of the resonator \cite{boissonneault2009dispersive}, and hence qualify as QND measurements. The scale where the QND behavior of the readout breaks down is quantified by the critical number of photons $n_{\text{crit}} = \Delta^2/4g^2$.}
 
  \textcolor{black}{Since the optimizer is based on the Hamiltonian \eqref{eq:Hfull}, the simulation is not necessarily limited to the dispersive regime. In principle, this allows the optimization to employ high readout power without concerns about the validity of the dispersive approximation. However, large power endangers the QND nature as the dispersive regime breaks down \cite{boissonneault2009dispersive}. The previous optimization target should help in that regard as it minimizes the number of photons in the cavity. To further ensure the QND nature of the readout, we add a QND-dedicated optimization target.}
 
 \subsubsection{Cost Functions}
\begin{figure}
  \centering \includegraphics[width=\columnwidth]{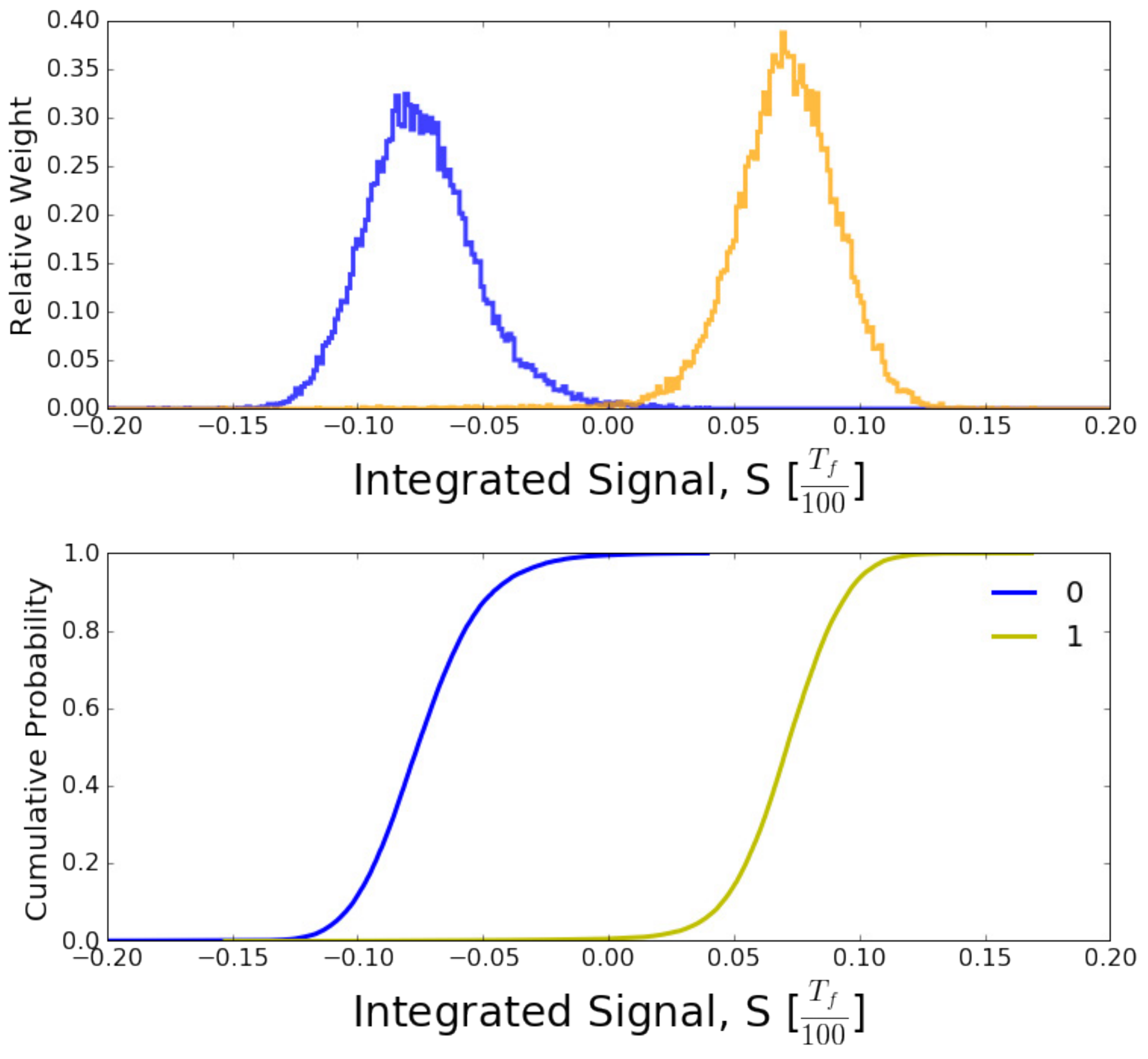}
  \caption{(a) Example of integrated readout signals for a qubit starting in the ground or excited state. The overlap between the two Gaussians contributes to the readout infidelity. One way to minimize the overlap between the distributions is by increasing the difference between the two Gaussian means.  (b) The corresponding cumulative probabilities of both distributions. The decision threshold that maximizes the fidelity is at the integrated signal value that maximizes the difference between the two cumulative probabilities.
  \label{figure:thresholds}}
\end{figure}

We include three cost functions for achieving the three optimization targets mentioned above simultaneously. Starting with the readout-fidelity cost function, we first inspect the process of post-measurement decision making. For every trajectory, the output signal $s(t) = \left\langle (a + a^{\dagger}) \right\rangle (t)$ is convoluted according to
 \begin{equation}
 \label{eq:filter}
 S = \int_0^{T_f} s(t) K(t) dt
 \end{equation}
 with a filter kernel $K(t)$ that is used to enhance the distinguishability of the readout signals \cite{gambetta2007protocols}. $T_f$ is the final measurement time. In order to determine the appropriate threshold value of $S$ distinguishing between readout of the qubit 0 and the 1 states, many trajectories are simulated and their corresponding values of $S$ are recorded. The histograms generated by the values of $S$ for both qubit states are fitted to Gaussians and the boundary is chosen to minimize the overlap between the two Gaussians. Histograms take on Gaussian form because of both the existence of noise in the measurement and the different evolution of each trajectory resulting in a distribution of integrated trajectory signals. Fig.\  \ref{figure:thresholds} shows an example of integrated readout signals. 
 
As Fig. \ref{figure:thresholds} suggests, one possible way to enhance readout fidelity is to increase the difference between the means of the two Gaussians to limit their overlap which is the main source of wrong decision making. Therefore, we implement the following fidelity cost function:
\begin{equation}
C_f = - \bigg(\frac{1}{T_f} \int_0^{T_f}  [\bar{s_0}(t) - \bar{s_1}(t)] \ dt\bigg)^2 .
\end{equation} 
Here, the bar denotes a trajectory average so that $\bar{s_i}(t)$ is the transmitted signal at time $t$  averaged over trajectories that all start in the initial state $\ket{i}$ with $i \in \{0,1\}$. 

\textcolor{black}{For the second optimization target, the average resonator occupation number at the end of the measurement needs to be minimized to enable fast resonator reset. In addition,  penalizing the average photon number during the entire measurement phase will make sure that dressed dephasing is suppressed. This also allows for potential termination of the measurement protocol at times smaller than the preset $T_f$ without accumulating large photon occupation.}  Hence, the second cost function was implemented in the form
\begin{equation}
C_r = \frac{1}{T_f}\sum_{i = 0,1} \int_0^{T_f} \overline{\left\langle ( a^{\dagger} a) \right\rangle_i} (t) dt.
\end{equation}
where $i \in \{0,1\}$ again refers to the initial state of the qubit.

Finally, for the QND optimization target, we add a cost function which rewards overlap between the final trajectory states and the corresponding initial states so that the qubit starting in the ground/excited state remains in the ground/excited state at the end of measurement.  As we always start the measurement with the resonator in the ground state, this cost function further helps reduce the cost $C_r$  as it ensures that the cavity returns back to the ground state. This QND cost function is taken to have the form
\begin{equation}
C_q = 1 - \overline{|\langle\psi_f|\psi_{i}\rangle|^2},
\end{equation}
where $|\psi_f\rangle$ and $|\psi_i\rangle$ are the final and initial states of each trajectory, respectively.
These three cost function are combined with other pulse-shaping constraints, and are assigned different weight factors which are set empirically by trial and error to improve convergence. 

\subsubsection{Implementation}
The problem is divided into two parts: optimization and classification. First, optimization is performed in such a way as to minimize the above-mentioned cost functions. To calculate the resulting readout fidelities, the diffusive trajectories produced by the resulting optimized pulse are mixed with Additive White Gaussian Noise of different powers, and then fed into an optimal-linear-filter classifier (see Appendix \ref{append:diffusive} for details).
The parameters used for the simulation are $\omega_q / 2 \pi = 4.6$\,GHz, $\omega_r / 2\pi = \omega_d / 2\pi = 5$\,GHz, $g/2\pi = 50$\,MHz, $\kappa = 50$\,M${ s}^{-1}$ and $\gamma = 1$ M${ s}^{-1}$.  These parameters correspond to $n_{\text{crit}} = 16$. We included 30 resonator levels and 3 transmon levels in the simulation. The simulated total time of measurement was taken to be $T_f = 100\,\text{ns} = 0.1\,T_1$. 

Following Ref.\ \cite{mcclure2016rapid}, we denote pulse amplitudes in dimensionless form $A_{n} = A/A_{\text{ph}}$. Here, $A$ is the absolute pulse amplitude, and $A_{\text{ph}}$ is the reference amplitude which results in a steady-state resonator occupation of one photon.  $n$ refers to the effective number of photons resulting from the used amplitude. 
We limit the pulse maximum amplitude to be $A_{16} = A_{n_{\text{crit}}}$. The fidelity of the optimized pulse is compared against the fidelity of a constant square pulse of amplitude $A_{n_{\text{crit}}}$. 

\subsubsection{Results}
\begin{figure}
  \centering \includegraphics[width=\columnwidth]{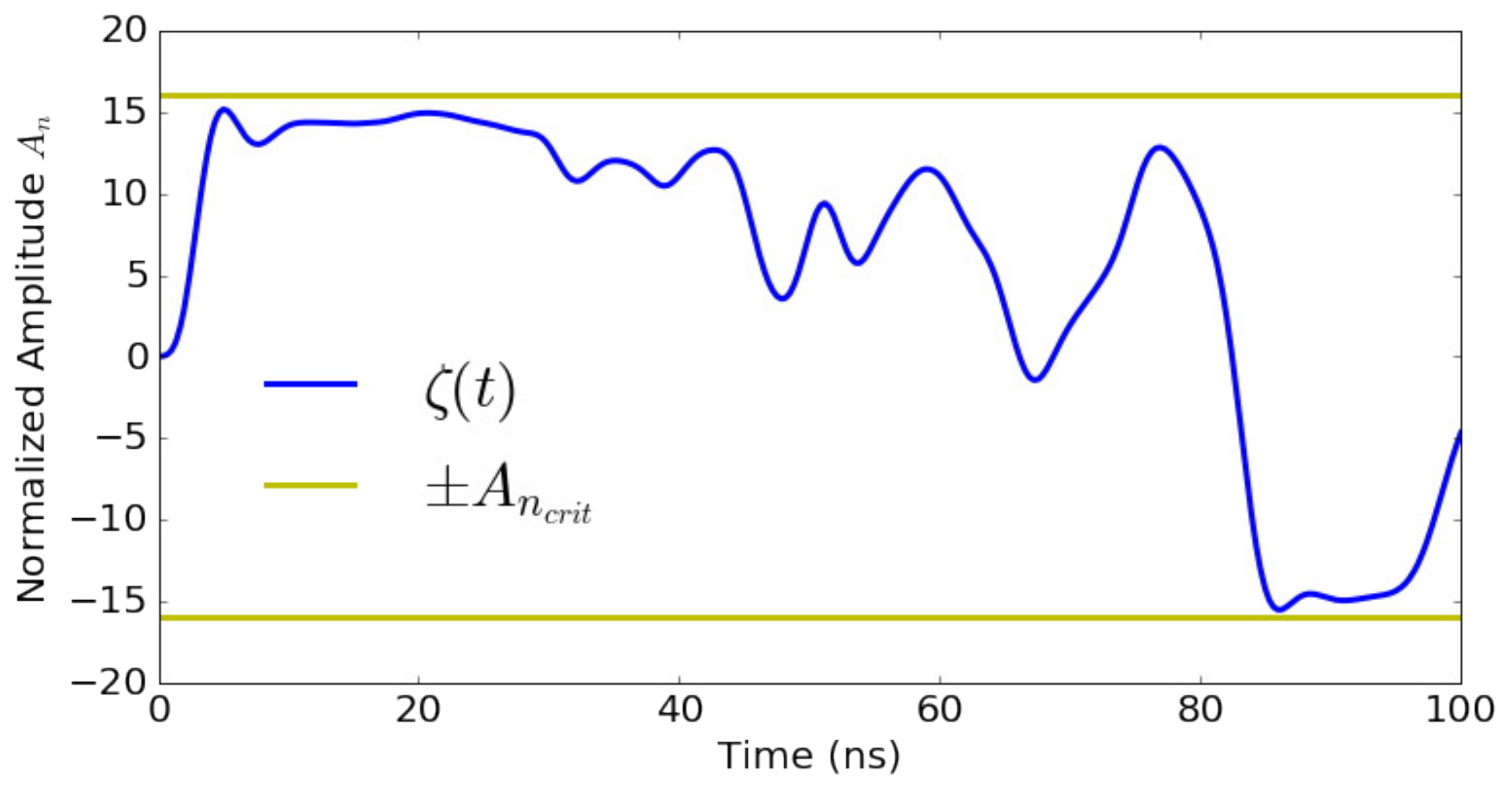}
  \caption{Optimized readout pulse within the amplitude limit. The pulse minimizes the weighted mixture of cost functions and uses only $58.3\%$ of the power needed by the constant pulse.
  \label{figure:readout}}
\end{figure}

\begin{figure}
  \centering \includegraphics[width=\columnwidth]{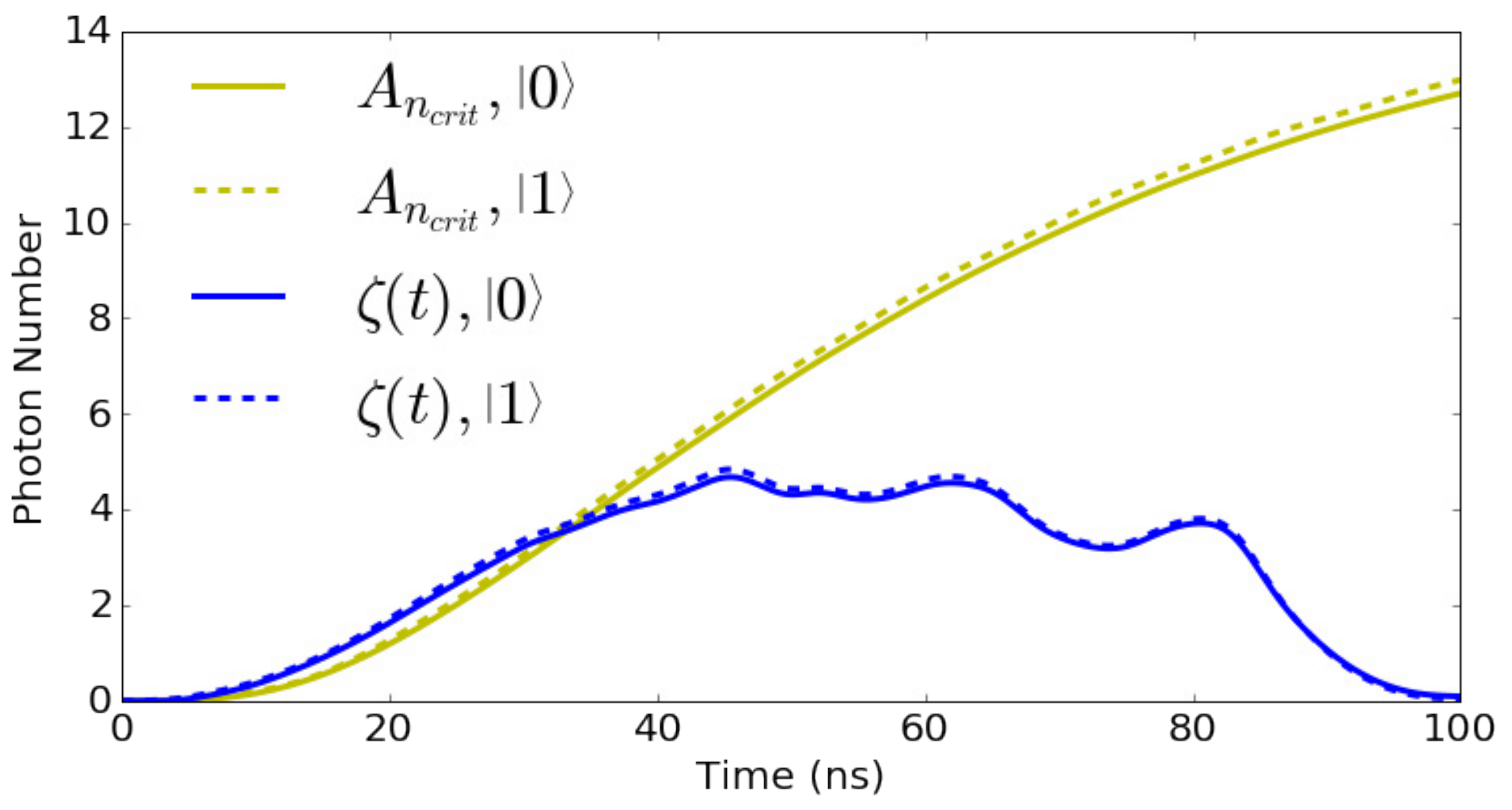}
  \caption{Resonator occupation as a function of measurement time for the optimized pulse $\zeta$(t) and the constant pulse of amplitude $A_{n_{\text{crit}}}$, with the qubit starting in the $\ket{0}$ and $\ket{1}$ states.
  \label{figure:photon}}
\end{figure}

We performed optimization with 8 parallel mini-batches, each of size $m_{\text{tot}} = 30$  using synchronous distributed training. With proper adjustments to the relative weights of different cost functions, the optimizer obtains the solution presented in Fig.\ \ref{figure:readout}. As ensured by the pulse-shaping cost functions, the pulse is smooth and starts and ends at near zero amplitudes.
The resulting resonator occupation is shown in Fig.\ \ref{figure:photon}.

As evident from Fig.\ \ref{figure:photon}, the optimized pulse maintains a relatively low photon number during the measurement process, and significantly decreases the photon numbers towards the end of the pulse. The optimized pulse achieves final photon numbers of 0.09 and 0.04 for the qubit starting in the $\ket{0}$ and $\ket{1}$ states; respectively. The constant pulse, by contrast, leads to occupations of around 13 for both states, which is two orders of magnitude higher than the optimized results. Therefore, optimization significantly improves the time needed for resetting the resonator after the measurement with the resonator almost empty already. If the active reset protocol proposed in Ref.\ \cite{boutin2017resonator} is used, then the reset is going to be 25 times faster than the non-optimized pulse case if the qubit is measured in the ground state, and 43 times faster if measured in the excited state.

\begin{figure}
  \centering \includegraphics[width=\columnwidth]{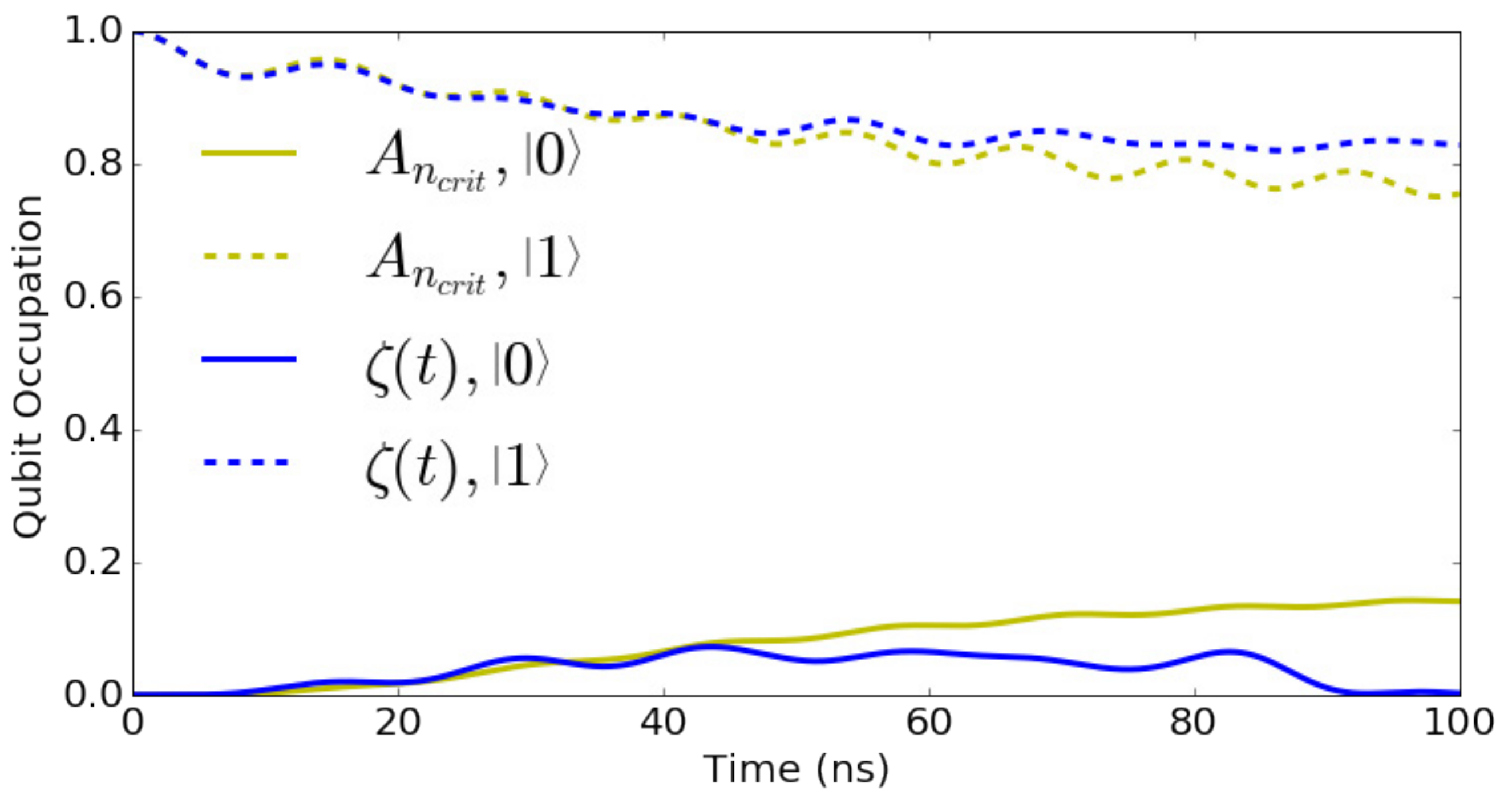}
  \caption{The qubit average occupation as a function of measurement time for both the optimized pulse $\zeta$(t) and the constant pulse of amplitude $A_{n_{\text{crit}}}$ with the qubit starting in the $\ket{0}$ and $\ket{1}$ states.
  \label{figure:qubit}}
\end{figure}

Moreover, to illustrate how optimization affects the QND nature of the readout, the qubit occupation numbers are plotted in Fig.\ \ref{figure:qubit} for both pulses. The results show that if one were to use a constant-power readout pulse, then the QND behavior would be compromised since the ground state gets excited to an average occupation of 0.14. The optimized pulse, however, manages to bring this occupation down to 0.002, ensuring the ground state measurement process to be QND within $99.8\%$. As for the excited state, the QND nature is limited by relaxation processes. Due to relaxation with the specified rate $\gamma$, the qubit occupation number would decay to around $0.9$ during the readout. However, due to potential qubit-cavity dressing, there are additional decay channels that may lower the final ideal occupation further. For instance, Ref.\ \cite{boissonneault2009dispersive} suggests that within the dispersive regime, the photon occupation of the resonator acts like an extra heat bath for the qubit. Fig.\ \ref{figure:qubit} shows that the constant pulse yields final occupation of $75.5\%$ while the optimized pulse increases it to $82.8\%$, thus enhancing the QND character of the measurement protocol.

\begin{figure}
  \centering \includegraphics[width=\columnwidth]{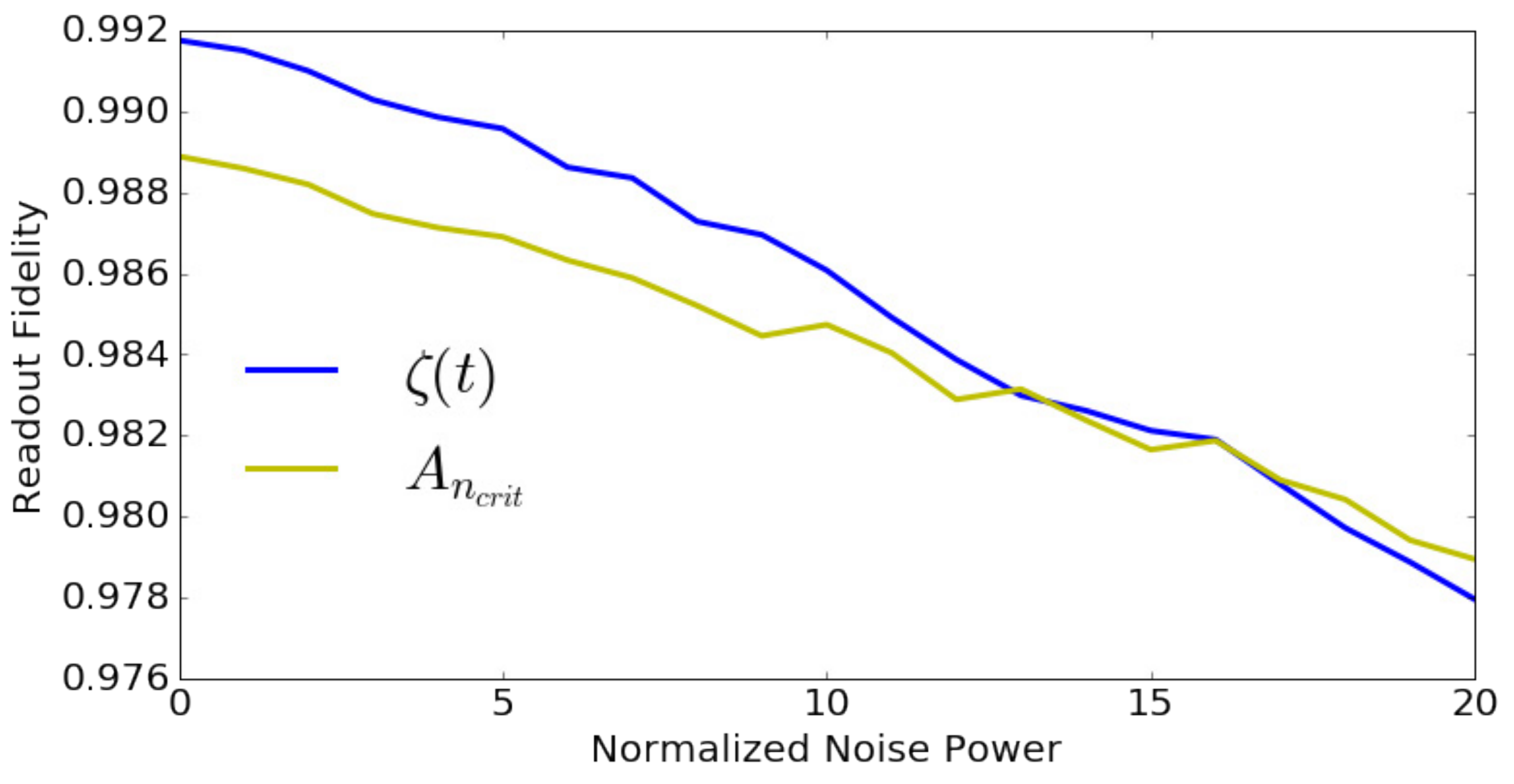}
  \caption{Readout fidelity of the optimized pulse compared to the constant, maximum-amplitude pulse.
  \label{figure:fidelity}}
\end{figure}

The QND and low-photon-number constraints ensure that the optimized pulse power is reduced from the maximum allowed power. Lowering the readout power could potentially cause the readout fidelity to decrease because of the presence of noise. To inspect whether readout power reduction negatively impacted the readout fidelities, we calculated the readout fidelities of the two pulses given different values for the noise power. The simulated noise power is again normalized in the same manner as the readout power, and we included normalized noise powers of up to $P_{20}$. The corresponding readout fidelities are presented in Fig.\ \ref{figure:fidelity}. The fidelities we obtain from both pulses are actually very close despite using almost half the total power. For most noise powers, the optimized pulse gives better fidelities. Therefore, we see that maintaining the QND nature of the readout and low photon-occupation numbers was achieved in a way that does not harm the readout fidelity.

In summary, our implementation allows for optimizing complex open systems under several constraints without having to analytically calculate their gradients. In addition, the simulation complexity is significantly reduced. For example, in this application, the Hilbert space dimension is 90. If traditional open-GRAPE methods were used, this would require propagating superoperators of dimension $8,100\times8,100$ . Instead, batches of size 90 by 30 (bigger or smaller batch sizes could also be used) are processed either in series or in parallel and lead to convergence with careful adjustment of convergence parameters such as learning rate, initial guess and relative cost function weights. The results suggest that readout could be performed with optimized pulses to control several aspects of the measurement. The example we presented suggests that the readout fidelities are not necessarily hurt by using smaller integrated powers, while significant benefits could be gained by allowing for constantly and smoothly changing amplitude pulses that restore the QND nature of the measurement and also allow for much faster resonator reset. 

\section{Conclusion}
In conclusion, we have harnessed the concept of automatic differentiation to build a flexible optimizer for open quantum systems. The optimizer is based on quantum trajectories, leading to significant reduction in the computational overhead compared to approaches based on density matrices. Combinations of improved-sampling techniques, generating mini-samples of trajectories for stochastic gradient descent and parallelization of trajectories are used according to the application to take advantage of the quantum trajectories nature of the optimizer. 

The optimizer was then utilized for both small and moderately sized quantum systems with quantum jumps being rare or common, and showed quick convergence to results that enhance over existing protocols. The optimizer could be used in the future to find optimal ways to control protected qubits such as the heavy fluxonium and 0-$\pi$ qubits which have similar structure to the $\Lambda$ system considered in the paper, but involve additional quantum levels and matrix elements. In this case, improved sampling will be very efficient as the jump probability during the gate time will generally be small, which allows for treating the system at a significantly reduced complexity. In addition, extra realistic optimization targets could be added to the employed cost functions. For example, control over the final gate time and the bandwidth of the used pulses are potential targets to include in the future. 

\section*{Acknowledgements}
We thank Nelson Leung for valuable discussions and for maintaining the hardware used in this research. We also acknowledge the support of NVIDIA\textsuperscript{\textregistered} Corporation through the donation of the Tesla\textsuperscript{\textregistered} K40 GPU used for this research. This research
was supported by the Army Research Office through Grant
No. W911NF-15-1-0421.

\appendix
\section{Automatic Differentiation}
\label{append:ad}
Here, we give an overview of the concept of automatic differentiation and examples of its usage. Automatic differentiation is a widely used concept in machine learning to generate the gradients of complex neural networks at runtime. An automatic differentiator allows users to define any mathematical relationships between an output function and its inputs, then can calculate the corresponding gradients by concatenating a series of pre-defined gradients. The full procedure of automatic differentiation is described in the following steps: 

\begin{enumerate}
\item The differentiator uses a basic set of $m$ pre-defined operations $\{f_i(x_1,x_2, \dots, x_{n_i})\}$ where $i = 1, 2, \dots, m$. For each operation $f_i$, the gradients $\frac{\partial f_i}{\partial x_1},\ldots,\frac{\partial f_i}{\partial x_{n_i}}$ with respect to all $n_i$ inputs must be defined. 
\item The user sets up the optimization problem by defining a cost function that is expressed solely by the operations $f_i$. This step creates the so-called computational graph. 
\item The algorithm runs through the computational graph in forward direction, starting from certain values of the inputs and computing the resulting cost-function value. The corresponding gradients are calculated as every operation is executed, and are saved for processing.
\item After completion of the forward path, reverse-mode automatic differentiation is utilized to trace backwards all paths relating the cost function to the inputs. This involves properly summing the stored partial derivatives of each path, generated by a recursive chain rule.% using the automatically calculated derivatives of every operation.
\end{enumerate}

In the following, we discuss a simple example where all operations involved act on scalars. Consider automatic differentiation of the cost function $C(x_1,x_2) = 2x_1^2 + \exp(x_1 x_2)$ with respect to its optimization inputs $x_1$ and $x_2$. To have an automatic differentiator, a set of building block operations needs to be predefined, one example of this set is shown in Table \ref{table:ad_set}.

\begin{table}[ht]
\caption{Example set of scalar operations for automatic differentiation} % title of Table
\centering % used for centering table
\begin{ruledtabular}
\begin{tabular}{c c c} % centered columns (4 columns)
 Name&  Definition & Gradients  \\ [0.5ex] % inserts table
%heading
\hline
MUL & $f(x_1,x_2) = x_1 x_2$ & $\frac{\partial f}{\partial x_1} = x_2$ and $\frac{\partial f}{\partial x_2} = x_1$ \\ [1ex] % inserting body of the table
DIV & $f(x_1,x_2) = \frac{x_1}{x_2}$ & $\frac{\partial f}{\partial x_1} = \frac{1}{x_2}$ and $\frac{\partial f}{\partial x_2} = \frac{-x_1}{x_2^2}$ \\ [1ex]
ADD & $f(x_1,x_2) = x_1 + x_2$ & $\frac{\partial f}{\partial x_1} = 1$ and $\frac{\partial f}{\partial x_2} = 1$ \\ [1ex]
SCALE & $f(a,x_1) = a\, x_1$ & $\frac{\partial f}{\partial x_1} = a$ \\ [1ex]
EXP & $f(x_1) = e^{x_1}$ & $\frac{\partial f}{\partial x_1} = e^{x_1}$ \\ [1ex]
\end{tabular}
\end{ruledtabular}
\label{table:ad_set} % is used to refer this table in the text
\end{table}

Note that the exponential operation does not strictly have to be included, since it could be represented in a Taylor series involving only MUL, ADD and SCALE operations. The same thing applies to other functions like  sine and cosine, as well as taking the integer power of an input, etc. In practice though, it is convenient to include special functions with compact analytical derivative in the basic set to give the user more flexibility in defining the computational graph. 

\begin{figure} 
  \centering \includegraphics[width=1\linewidth]{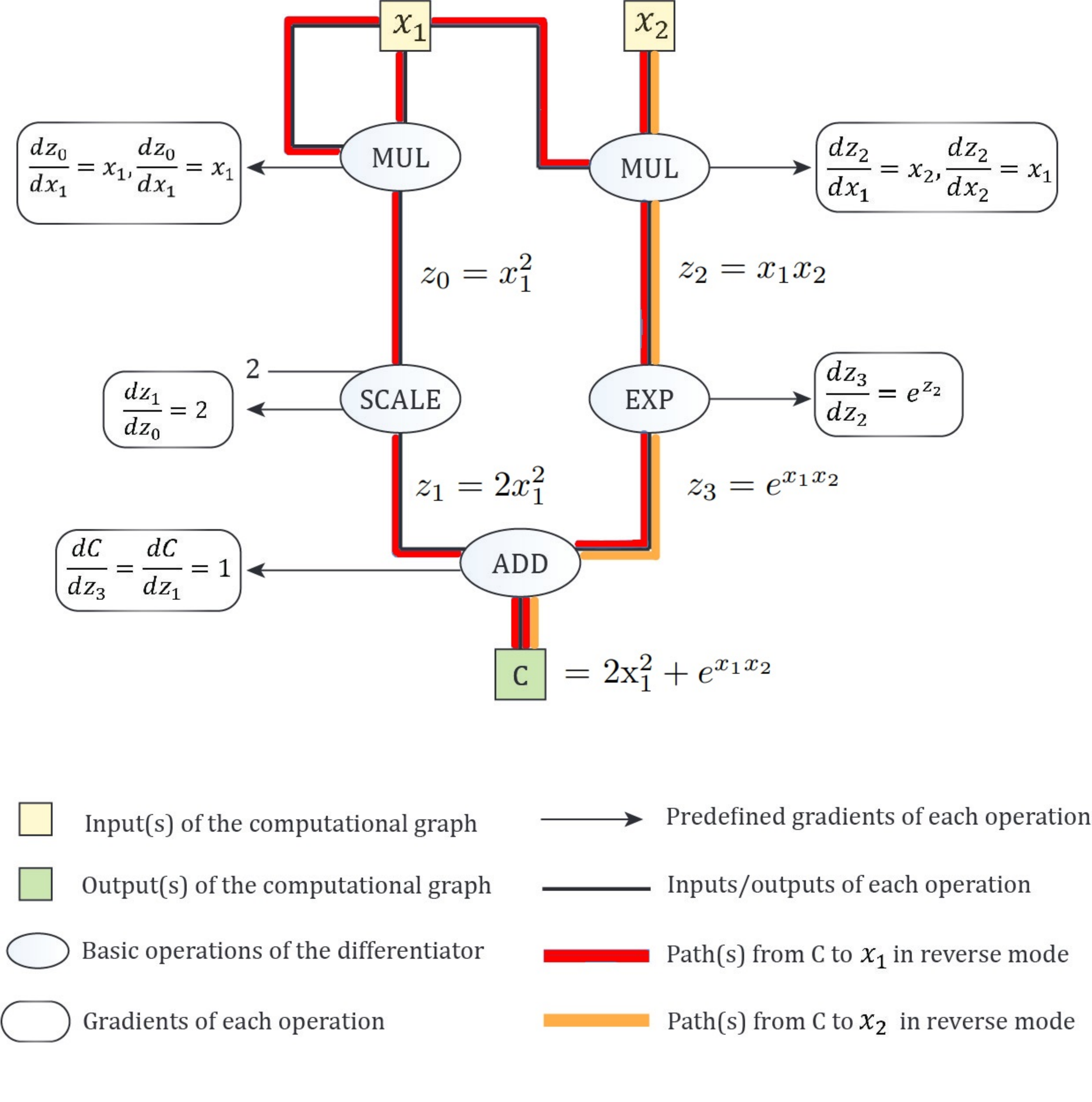}
  \caption{ Computational graph for the example cost function $C(x_1,x_2) = 2x_1^2 + \exp(x_1 x_2)$, created for automatic differentiation to calculate $\frac{\partial C}{\partial x_1}$ and $\frac{\partial C}{\partial x_2}$. Every ellipse represents one basic operation whose gradients are known  and symbolically given by the expressions in rounded rectangles, attached by arrows. The inputs to each operation are represented by lines entering the ellipse from above or from the left/right. The output of each operation is given a name $z_i$ written on the output line emerging from the bottom. After the computational graph run in forward direction (from $x_1$, $x_2$ towards $C$), reverse-mode automatic differentiation identifies paths between $C$ and each of $x_1$ and $x_2$. In this example, there is one path (orange color) relating $C$ to $x_2$ while there are 3 paths (red paths) between $C$ and $x_1$. Gradients are automatically calculated in a backward fashion by recursively multiplying the gradients in each path from bottom to top, then summing over all paths contributing to the same variable. 
  \label{figure:ex_comp_graph}}
\end{figure}

Using Table \ref{table:ad_set} as the basic-operation set, we next define the computational graph, see Fig.\ \ref{figure:ex_comp_graph}. After the computational graph is run in forward direction (from $x_1$ and $x_2$ towards $C$), reverse-mode automatic differentiation identifies paths between $C$ and each of $x_1$ and $x_2$. In this example, there is one path (the orange path) relating C to $x_2$ while there are 3 paths (red paths) between $C$ and $x_1$. Gradients are then automatically calculated in a backward fashion by recursively multiplying the gradients in each path from bottom to top, then summing over all paths contributing to the same variable. Following this scheme, automatic differentiation will calculate the gradients in the reverse mode to be 
\begin{align*}
\frac{\partial C}{\partial x_1} &= \frac{\partial C}{\partial z_1} \frac{\partial z_1}{\partial z_0} \frac{\partial z_0}{\partial x_1} + \frac{\partial C}{\partial z_1} \frac{\partial z_1}{\partial z_0} \frac{\partial z_0}{\partial x_1} + \frac{\partial C}{\partial z_3} \frac{\partial z_3}{\partial z_2} \frac{\partial z_2}{\partial x_1} \\ &=   1 \cdot 2 \cdot x_1 + 1 \cdot 2 \cdot x_1 + e^{z_2} x_2 = 4 x_1 + e^{x_1 x_2} x_2, \\ 
\frac{\partial C}{\partial x_2} &= \frac{\partial C}{\partial z_3} \frac{\partial z_3}{\partial z_2} \frac{\partial z_2}{\partial x_2} = 1 \cdot e^{z_2} x_1 = e^{x_1 x_2} x_1,
\end{align*}
which are indeed the correct gradients. In this way, automatic differentiation allows for calculating gradients for any computational graph that is defined in terms of the building blocks of the differentiator. 

To utilize automatic differentiation in the optimization of quantum simulations, the set of basic operations must include matrix operations since the quantum dynamics is propagated through matrix operations. The definition of gradients then needs to account for the non-commutativity of matrix multiplication. To be able to generate gradients in the backward paths, the differentiator must be able to express the gradients with respect to the outputs in terms of the gradients with respect to the inputs of the operation in hand. Thus, for every basic matrix operation $\boldsymbol{f}(\boldsymbol{M_i})$, %instead of just defining the gradient of the operation $\frac{\partial \boldsymbol{f}}{\partial \boldsymbol{M_i}}$, 
we need to specify how the gradients with respect to the inputs of the operation, $\frac{\partial C}{\partial \boldsymbol{M_i}}$, can be calculated given gradients with respect to the output $\frac{\partial C}{\partial \boldsymbol{f}}$.  Table \ref{table:matrix_set} shows some of the basic matrix operations and their gradient relations needed for automatic differentiation.

\begin{table}[ht]
\caption{Examples of basic matrix operations needed for automatic differentiation. Arguments of MATMUL must be compatible under matrix multiplication; arguments of ADD must have the same dimensions, and DET is only defined for square matrices.} % title of Table
\centering % used for centering table
\small
\begin{ruledtabular}
\begin{tabular}{c c c} % centered columns (4 columns)
 %inserts double horizontal lines
 Name&  Definition & Gradients  \\ [0.5ex] % inserts table
%heading
\hline  % inserts single horizontal line
MATMUL & $\boldsymbol{f}(\boldsymbol{M_1},\boldsymbol{M_2)} = $ & $\frac{\partial C}{\partial  \boldsymbol{M_1 }} =  \boldsymbol{ M_2} \frac{\partial C}{\partial  \boldsymbol{f}} $ \\ [1ex] % inserting body of the table
& $\boldsymbol{M_1 M_2}$& and $\frac{\partial C}{\partial \boldsymbol{M_2}} =  \frac{\partial C}{\partial f} \boldsymbol{M_1} $ \\[1ex]
ADD & $\boldsymbol{f}(\boldsymbol{M_1},\boldsymbol{M_2)} = $ & $\frac{\partial C}{\partial \boldsymbol{M_1}} =  \frac{\partial C}{\partial \boldsymbol{f}} $ \\ [1ex] % inserting body of the table
& $\boldsymbol{M_1} + \boldsymbol{M_2}$& and $\frac{\partial C}{\partial \boldsymbol{M_2}} =  \frac{\partial C}{\partial \boldsymbol{f}}  $ \\[1ex]
SCALE & $\boldsymbol{f}(a,\boldsymbol{M_1}) = a \boldsymbol{M_1}$ & $\frac{\partial C}{\partial \boldsymbol{M_1}} =  a \frac{\partial C}{\partial \boldsymbol{f}}  $ \\ [1ex]
TRACE  & $f(\boldsymbol{M_1}) = \Tr(\boldsymbol{M_1})$ & $\frac{\partial C}{\partial \boldsymbol{M_1}} =  \boldsymbol{\mathbb{1}} \frac{\partial C}{\partial f}  $ \\ [1ex]
DET  & $f(\boldsymbol{M_1}) = \det(\boldsymbol{M_1})$ & $\frac{\partial C}{\partial \boldsymbol{M_1}} = \det(\boldsymbol{M_1}) \boldsymbol{M_1}^{-T} \frac{\partial C}{\partial f}  $ \\ [1ex]
TRANSPOSE  & $\boldsymbol{f}(\boldsymbol{M_1}) = \boldsymbol{M_1}^{T}$ & $\frac{\partial C}{\partial \boldsymbol{M_1}} =   \frac{\partial C}{\partial \boldsymbol{f}}  $ \\ [1ex] 
CONJUGATE  & $\boldsymbol{f}(\boldsymbol{M_1}) = \boldsymbol{M_1}^{*}$ & $\frac{\partial C}{\partial \boldsymbol{M_1}} =   \frac{\partial C}{\partial \boldsymbol{f}}  $ \\ [1ex]
% [1ex] adds vertical space
 %inserts single line
\end{tabular}
\end{ruledtabular}
\label{table:matrix_set} % is used to refer this table in the text
\end{table}

Because of the asymmetry in the MATMUL-gradient rule, it is not simple to express the gradient relations for most special functions like it was in the scalar case. Most of the functions not defined in Table \ref{table:matrix_set} (including the matrix exponential) must be expressed in terms of the basic operations like MATMUL, ADD and SCALE so that the gradient calculation is accurate. 

\section{Analytical Gradient of State Transfer in Quantum Trajectories}
In this appendix, we will show how the analytical gradients of the cost function 
\begin{equation}
C = 1 - |\langle\psi_T|\psi_{N}\rangle|^2
\end{equation}
could be calculated in a quantum trajectory.
\label{append:analytical}
The final state in a quantum trajectory can be written as 
\begin{equation}
%\begin{split}
\ket{\psi_N} = \frac {M_N M_{N-1} \cdots M_1 \ket{\psi(0)}} {\norm{M_N M_{N-1} \cdots M_1 \ket{\psi(0)}}}
%\end{split}
\end{equation}
where the non-unitary propagators $M_j$ are given by
\begin{equation}\label{Ms1}
M_j = 
\exp(\textstyle -i H_j dt - \frac{dt}{2} \sum_l \gamma_l c_l^{\dagger} c_l )
\end{equation}
in the absence of a jump, or
\begin{equation}\label{Ms2}
M_j = c_l,
\end{equation}
if a jump occurs in decoherence channel $l$.
If we define \begin{equation}
F_j = \sqrt{ \bra{\psi(0)} M_1^{\dagger} \cdots  M_j^{\dagger} M_j  \cdots M_1 \ket{\psi(0)}}
\end{equation}
and $Y=F^2$ with $F=F_N$, then the gradient in the case of no jump is
\begin{equation}
\begin{split}
&\frac{\partial Y}{\partial u_{kj}} = \\ &\frac{\partial}{\partial u_{kj}} \bra{\psi(0)} M_1^{\dagger} \dots M_{N-1}^{\dagger} M_N^{\dagger} M_N M_{N-1} \dots M_1 \ket{\psi(0)} = \\ & i dt \bra{\psi(0)} M_1^{\dagger} \dots  M_j^{\dagger} H_k \dots M_N^{\dagger} M_N M_{N-1} \dots M_1 \ket{\psi(0)} + \\ & 
 - i dt \bra{\psi(0)} M_1^{\dagger}  \dots M_N^{\dagger} M_N  \dots  H_k M_j  \dots M_1 \ket{\psi(0)} = \\& + 2 dt  F F_j \iim ( \bra{\psi_N}     M_N\dots M_{j+1}H_k\ket{\psi_j})
\end{split}
\end{equation}

Hence, \begin{equation}
\frac{\partial (\frac{1}{F})}{\partial u_{kj}} = \frac{\partial (\frac{1}{F})}{\partial Y}\frac{\partial Y}{\partial u_{kj}} = -\frac{dt  F_j \iim ( \bra{\psi_N}     M_N\dots M_{j+1}H_k\ket{\psi_j})}{F^2}
\end{equation}
Therefore, we can finally write the dependence of the final state $\ket{\psi_N}$ on the control parameters as

\begin{equation}
\begin{split}
\frac{\partial \ket{\psi_N}}{\partial u_{kj}} = \\ &\frac{- i dt F_j M_N M_{N-1} \dots M_{j+1} H_k \ket{\psi_j}}{F} \\ &- \frac{dt  F_j \iim ( \bra{\psi_N}     M_N\dots M_{j+1}H_k\ket{\psi_j}) \ket{\psi_N}}{F}
\end{split}
\end{equation}
Then, if no jump happens at time step j, 
\begin{widetext}
\begin{equation}
\begin{split}
\frac{\partial C}{\partial u_{kj}} &=  - \bra{\psi_T} \frac{\partial \ket{\psi_N}}{\partial u_{kj}} \langle \psi_N | \psi_{T}\rangle - \langle \psi_T | \psi_{N}\rangle \frac{\partial \bra{\psi_N}}{\partial u_{kj}} \ket{\psi_T} \\ &= - \langle \psi_N | \psi_{T}\rangle \bra{\psi_T} \big[ \frac{- i dt F_j M_N M_{N-1} \dots M_{j+1} H_k \ket{\psi_j}}{F}  - \frac{dt  F_j \iim ( \bra{\psi_N}     M_N\dots M_{j+1}H_k\ket{\psi_j}) \ket{\psi_N}}{F}\big] \\ &- \langle \psi_T | \psi_{N} \rangle \big[\frac{ i dt F_j \bra{\psi_j} H_k M_{j+1}^{\dagger} \dots  M_{N-1}^{\dagger} M_{N}^{\dagger} }{F} - \frac{dt  F_j \iim ( \bra{\psi_N}     M_N\dots M_{j+1}H_k\ket{\psi_j}) \bra{\psi_N}}{F}\big] \ket{\psi_T} \\ &= 
\frac{-2 F_j dt}{F}  \iim \bigg [ 
\langle \psi_T| \Big[  \prod_{j'>j} M_{j'} \Big]   
 H_k  | \psi_{j}\rangle \langle \psi_N | \psi_T \rangle \bigg ]+ (C-1) \frac{dt  F_j \iim ( \bra{\psi_N}    \Big[  \prod_{j'>j} M_{j'} \Big] H_k\ket{\psi_j}) }{F}
\end{split}
\end{equation}
\end{widetext}
If we define $f_{jk}(\psi) = \bra{\psi}    \Big[  \prod_{j'>j} M_{j'} \Big] H_k\ket{\psi_j} $, then the analytical gradients take the form 
\begin{equation}
\label{actual grads}
\frac{\partial C}{\partial u_{kj}} = \begin{cases}

\frac{-2 F_j dt}{F}  \iim ( f_{jk}(\psi_T)\langle \psi_N | \psi_{T}\rangle ) & \\ + \frac{(C-1)}{F} dt  F_j \iim ( f_{jk}(\psi_N)) &\quad\text{no jump} \\
0  &\quad\text{jump}
\end{cases}
\end{equation}
\normalsize
Note that the extra gradient term due to the decay of the norm of the state is only negligible at the beginning of training when $C \approx 1$ but as the algorithm enhances the fidelity, this term becomes more and more dominant.

This whole analysis also assumed that a jump occupies a whole time step while in principle it should not. Instead, a more accurate approach would be to integrate the state to find the time $t_{\text{jump}}$ when its norm reaches the random number $r$, apply a quantum jump at that time, and then keep propagating/integrating the resulting state starting again from $t_{\text{jump}}$. In this case, the corresponding analysis to find the analytical gradient would be even more complex since it clearly deviates from the standard picture of having one propagator at every time step. 

\section{Readout Fidelity Implementation}
\label{append:diffusive}
In this appendix, we detail our implementation of the calculation of the readout fidelity  as defined in Eq.\ \eqref{eq:fid}. The qubit-cavity readout application in section \ref{sec:readout} is of a different nature than the rest of applications, since it does not include the usual penalization of a state-transfer fidelity. Instead, several targets are desired to be minimized as specified by the cost functions $C_f, C_r$ and $C_q$. We note that the readout fidelity itself is not one of the cost functions used in the simulation. 

In order to keep the same structure of the optimizer code from the previous applications, we separated the simulation into two parts. First, the optimization part is utilized to minimize the three cost functions according to their defined relative weights. Then a classifier code is used afterwards to calculate the readout fidelity given the resulting optimized pulse. The classifier generates a sample of trajectories starting from the ground and excited states, then adds random noise of different powers to each trajectory and finally calculates the resulting fidelities. 

First, for the trajectories generation part, the classifier uses the optimized pulse to simulate the dynamics of the stochastic Schr{\" o}dinger equation (SSE) that represents diffusive trajectories of the qubit-resonator system. Since  readout is implemented in the lab via a homodyne measurement of the resonator \cite{schuster2007circuit}, the resulting trajectories are of diffusive (not jump) type. The optimizer still uses jump trajectories like all the other applications, which is allowed since all the used cost functions only involve averages over trajectories. Regardless of the type of unraveling used (e.g. jump vs diffusive), the averaged expectation values of operators remain the same, which match the results from the master equation. However, for the readout-fidelity calculation, it is crucial to simulate the actual diffusive trajectories that are observed in the lab since the classification is a process that depends on how the individual trajectories could be distinguished.

The SSE used to generate the correct diffusive trajectories is \cite{armen2006low,van2005quantum,Wiseman_Milburn_2009}:
\begin{align}
&d \ket{\psi} = \big[\sqrt{\kappa} a \ dW_1 +  \sqrt{\gamma} b \ dW_2 +  \\ \nonumber
&dt (- i H - \frac{\kappa}{2}  a^{\dagger}a  - \frac{\gamma}{2}  b^{\dagger}b + \kappa a  \left\langle  a^{\dagger} + a \right\rangle + \gamma b  \left\langle  b^{\dagger} + b \right\rangle)  \big] \ket{\psi}
\end{align}
where $dW_1$ and $dW_2$ are independent Wiener increments that satisfy  $E(dW) = 0$ and $E ((dW)^2) = dt$. ($a$, $\kappa$) and ($b$, $\gamma$) are the lowering operators and decay rates for the resonator and qubit, respectively. Solutions of this SSE were obtained using the Euler-Maruyama method with Richardson extrapolation \cite{kloeden2011numerical}.

After obtaining the trajectories, noise is added with different powers to simulate multiple sources of experimental readout noise. The noisy trajectories are then multiplied by an optimal linear filter \cite{gambetta2007protocols} and integrated over the measurement time as in Eq. \eqref{eq:filter}. The classifier finally decides on a threshold value for the integrated signals that maximizes the fidelity and calculates the corresponding value of the fidelity.

\end{document}